\documentclass{article}
\usepackage{a4wide}
\usepackage{pdflscape}
\usepackage{graphicx}
\usepackage{amsmath}
\usepackage{amssymb}
\usepackage{lipsum} %lorem ipsum
\usepackage{mathpartir}
\usepackage{mathtools}
\usepackage{arydshln}
\usepackage{multirow, multicol} %advanced tabular features 
\usepackage{rotating}
\usepackage{wrapfig}
\usepackage{subcaption}
\usepackage{caption}
\usepackage{longtable}
\usepackage{xcolor}
\usepackage{tikz}
\usepackage{pgffor}
\usetikzlibrary{arrows.meta, fit, automata, positioning, calc, trees, shapes.multipart}
\usepackage{pgf-umlsd}
\usepackage{xspace}
\usepackage{orcidlink}
\usepackage{titling}

\newtheorem{definition}{Definition}[section]

\providecommand{\keywords}[1]
{
  \small	
  \textbf{\textit{Keywords---}} #1
}

\renewcommand{\mess}[4][0]{
  \stepcounter{seqlevel}
 \path
  (#2)+(0,-\theseqlevel*\unitfactor-0.7*\unitfactor) node (mess from) {};
  \addtocounter{seqlevel}{#1}
  \path
  (#4)+(0,-\theseqlevel*\unitfactor-0.7*\unitfactor) node (mess to) {};
  \draw[->,>=angle 60] (mess from) -- (mess to) node[midway, above]
  {#3};    
}

\newcommand{\cpp}{\texttt{c++}\xspace}

\newcommand{\st}{\textit{st}}
\newcommand{\stglob}{\sigma_g}
\newcommand{\stloc}{\sigma_l}
\newcommand{\ID}{\textit{ID}}
\newcommand{\semE}{\text{E\!\!\!\!E}}

\newcommand{\tab}[2]{\setlength{\hspace*}{0.4cm * #2}#1\setlength{\hspace*}{0.4cm - \widthof{#1}}}
\newcommand{\STF}[5]{\textbf{frame}(#1, #2, #3, #4) :: #5}
\newcommand{\pid}{\textit{id}_p}
\newcommand{\fid}{\textit{id}_f}

\newcommand{\TrS}[1]{\mathbb{T}_S( #1, l)}
\newcommand{\TrE}[1]{\mathbb{T}_E( #1)}
\newcommand{\cat}{+\!\!+}
\newcommand{\Spre}[1]{S_{\textit{pre}_{#1}}}

\newcommand{\whr}{\textbf{where}~}
\newcommand{\emptylist}{[]}
\newcommand{\State}{\textit{ID}_x \to V}

\newcommand{\tlprog}{\textit{tl}(\textit{prog})}

\title{Semi-Automatic Extraction of Formal Models from Object Oriented Code
\thanks{This publication is part of the PVSR project (with project number 17933) of the MasCot research programme which is financed by the Dutch Research Council (NWO).} \vspace{0.3ex}\\
\small(using the novel SSTraGen tool)}
\author{P.H.M.\ van Spaendonck \orcidlink{0000-0002-9536-1524}}

\date{}

\begin{document}
\maketitle
\begin{abstract}
Behavioral models are incredibly useful for understanding and validating software.
However, the automatic extraction of such models from actual industrial code remains a largely unsolved problem with current solutions often not scaling well with the complexity and size of industrial systems or having to rely on approximations.
To enable the extraction of useful models from code, we provide a framework for transforming object-oriented code into processes from which, when paired with minimal user input, models can be automatically generated and composed.
Paired with this, we introduce the novel SSTraGen (StateSpace Transformation \& Generation) tool, which provides an implementation of this framework.
Through case studies at Philips Image Guided Therapy Systems, we showcase the practical applicability and usefulness of this tool, including the transformation of a component with $>1000$ LOC.
\end{abstract}

\section{introduction}
Behavioral models of software have, time and time again, shown to be incredibly useful when it comes to understanding and maintaining complex cyber-physical systems.
For example, models can be used to skip the need to manually write tests via model-based testing \cite{tretmans2008model,utting2010practical}, verification of important behavioral requirements can be done by solving modal formulae \cite{baier2008principles,hennessy1980observing,mateescu2003efficient}, finding deadlocks and race conditions in complex asynchronous systems becomes almost trivial \cite{10.1145/503271.503216}, and proving equivalence between multiple models has shown to be an effective approach towards proving the correctness of a parallel and communicating systems \cite{slidingwindow,FREDLUND1997459,Shankland}.

However, in practice, acquiring and maintaining such models is often the hard part.
Doing so manually is often too complex and labor-intensive.
Similarly, the automatic extraction/reverse engineering of such models from industrial code comes with its own caveats which we discuss in the Related Work section.
This forms a significant obstacle to the integration of model-based techniques in the industry.

\begin{figure}
    \centering
    \includegraphics[width=.95\linewidth]{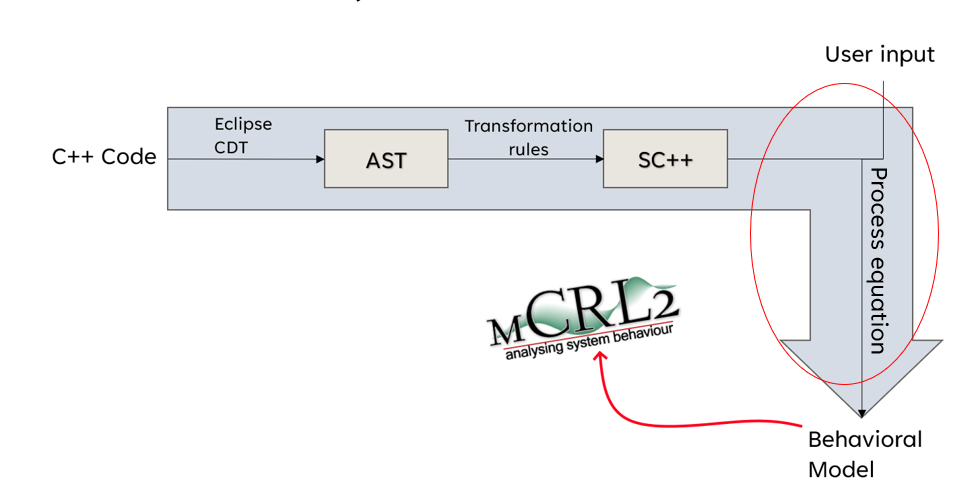}
    \caption{Framework for transforming object-oriented code into behavioral models}
    \label{fig:framework}
\end{figure}

We outline a semi-automated model extraction framework, see Figure \ref{fig:framework}, for transforming object-oriented code into proper scalable behavioral models.
The framework consists of a three-step process:
First, the code is parsed into an AST.
Second, we apply transformation rules specified by us on the AST, and generate an intermediate data structure.
Third, this data structure, paired with user input to bound possible infinite behavior, is then used to calculate the behavioral model, by transforming it into a process equation, which effectively acts as the behavioral model of our code.

The behavioral models obtained through this framework capture all of the possible externally observable behavior of a given class instance, i.e.\ method calls and field accesses to and from the transformed instance, under run-to-completion semantics.
When considering multiple simultaneous classes, e.g.\ members and instances with bounded lifespan, these instances are modeled, either through the same framework or manually, and the parallel composition of all individual models is then used to calculate the total model of the combined system.
These models can be acquired in the same way as the main class, or replaced with simple abstractions.
We note that the latter is very useful, if not crucial, when considering complex systems, as it allows us to circumvent the need to transform the entire system down to the level of the \cpp standard library.

\begin{figure}
\centering
    \begin{subfigure}[b]{0.42\textwidth}
    \centering
    \includegraphics[width=\linewidth]{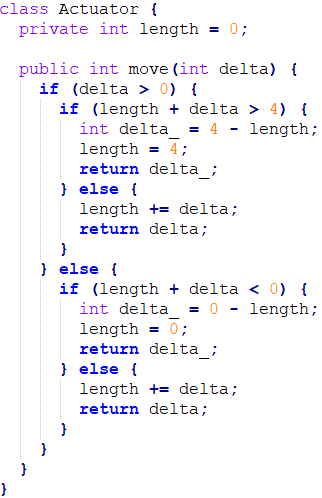}
    \caption{\label{fig:pseudocode_actuator}}
    \end{subfigure}
\quad
    \begin{subfigure}[b]{0.42\textwidth}
    \centering
    \begin{tikzpicture}
    \tikzstyle{every node}=[font=\scriptsize]
    \node[circle, draw] (l0decr) at (0, .9cm) {};
    \foreach \i [evaluate=\i as \y using \i*-1.5,
    evaluate=\i as \ni using int(\i + 1)] in {0,1,2,3} {
        \node[circle, draw, fill=lightgray] (l\i) at (0 cm,\y cm) {};
        \node[circle, draw] (l\i incr) at (.6cm, \y cm -.8cm) {};
        \node[circle, draw] (l\ni decr) at (-.6cm, \y cm -.8cm) {};
    }
    \node[circle, draw, fill=lightgray] (l4) at (0, -6cm) {};
    \node[circle, draw] (l4incr) at (0, -6.9cm) {};

    %Top (l0decr
    \draw[->] (l0) -- ++(-.3cm, .1cm) -- node[pos=.5, left]{\texttt{call(move, [-1])}} ++(0cm, .7cm) -- (l0decr);
    \draw[<-] (l0) -- ++(.3cm, .1cm) -- node[pos=.5, right]{\texttt{return(move, 0)}} ++(0cm, .7cm) -- (l0decr);

    \foreach \i [evaluate=\i as \ni using int(\i + 1)] in {0,1,2,3} {
        \draw[->] (l\i) -- ++(.6cm,-.2cm) -- (l\i incr)
            node[pos=0.5, right]{\texttt{call(move, [1])}};
        \draw[->] (l\i incr) -- ++(0,-.6cm) -- (l\ni)
            node[at start, right]{\texttt{return(move, 1)}};

        \draw[<-] (l\i) -- ++(-.6cm,-.2cm) -- (l\ni decr)
            node[pos=0.5, left]{\texttt{return(move, -1)}};
        \draw[<-] (l\ni decr) -- ++(0,-.6cm) -- (l\ni)
            node[at start, left]{\texttt{call(move, [-1])}};
    }

    %Bottom (l4incr)
    \draw[<-] (l4) -- ++(-.3cm, -.1cm) -- node[pos=.5, left]{\texttt{return(move, 0)}} ++(0cm, -.7cm) -- (l4incr);
    \draw[->] (l4) -- ++(.3cm, -.1cm) -- node[pos=.5, right]{\texttt{call(move, [1])}} ++(0cm, -.7cm) -- (l4incr);
    
    \end{tikzpicture}
    \caption{\label{fig:actuator_model}}
    \end{subfigure}
\caption{(\subref{fig:pseudocode_actuator}) Actuator example code, (\subref{fig:actuator_model}) Model of the example code.}
\end{figure}

Let us consider the framework in more detail using the code shown in Figure \ref{fig:pseudocode_actuator}.
The given code describes a class, which is used to keep track of the length of a physical actuator.
The real world actuator should never be extended further than 4cm, or shrink below 0cm.
When the move function is called with a desired extension/reduction, the class will return the actual amount by which the real world actuator can be moved and update its length accordingly.
We're interested in acquiring a model of this class such that we can verify that it will never exceed the given bounds of 0 and 4cm.

To do so, the Actuator class is first transformed into the intermediate data structure as per the framework.
Second, we must bound the set of possible arguments with which the \texttt{mode} method can be called as considering all machine integers is impractical.
As such, we bound \texttt{delta} to only the integers $-1$, and $1$.
The intermediate data structure is now transformed into a process equation and fed through the model-based tool set mCRL2 \cite{mcrl2toolset} to generate the actual state space, and apply action-renaming and state space reduction to acquire a concise, yet equally expressive model, shown in Figure \ref{fig:actuator_model}, of the externally observable behavior of an instance of the Actuator class.

After the transformation process, we can see that the model has exactly 5 states, colored gray, in which a function can be called.
Each of these states corresponds to one of the possible lengths of the actuator.
Whenever the \texttt{move} function is called, we move to a state, colored white, in which the function is executed and eventually returns a value and we move to our possible new length.
From the model we can also see that whenever the value $0$ is returned, we always end up in the same state in which the \texttt{move} method was originally called.
With this model it is fairly easy to prove using model verification techniques that there is no sequence of \texttt{move} calls possible in which we end up with a length outside the safety bounds.

Paired with this transformation framework, we introduce the novel GUI based SSTraGen (StateSpace Transformation \& Generation) tool, which provides an implementation of this theory and guides the user in providing the correct/optimal input needed to complete the transformation from \cpp code to Behavioral model.
Through case studies at Philips Image Guided Therapy Systems, we showcase the practical applicability and usefulness of this tool and its underlying theory, including the transformation of a component with ${>}1000$ lines of code.

The paper is split up as follows:
In Section \ref{sec:related-work}, we discuss other known techniques for acquiring models from code.
In Section \ref{sec:framework}, we go over the entire transformation framework that transforms actual \cpp code to a behavioral model.
In Section \ref{sec:sstragen}, we discuss our implementation of this framework in the form of the novel SSTraGen tool.
In Section \ref{sec:case-study}, we discuss how the SSTraGen tool was utilized to acquire models of actual industrial code and provide insights into the behavior of some of the complex systems found within Philips IGT.
We conclude our work in Section \ref{sec:conclusion}, in which we also provide potential future work relating to the framework and tool shown in this paper.

\section{Related Work} \label{sec:related-work}
The scope of our research is not focused on the actual verification or analysis of software, but on the extraction of formal models, that capture at least part of the software behavior.
For these models, verification and analysis are only one of their many uses.
The type of model delivered by these techniques varies.
Generally speaking, model extraction techniques either deliver some behavioral UML diagram, e.g.\ use-case or sequence diagrams, or a state machine/labeled transition system.
We consider techniques that extract structural UML diagrams, e.g.\ class diagrams, to be outside the scope of this research, as these only contain the static aspects of a given system.

The most common type of behavioral UML diagram produced by current techniques, as is outlined in the literary overview in \cite{briand2004towards}, is the UML sequence diagram.
This type of diagram focuses on a particular sequence/execution, instead of capturing all possible executions of a given software program.
State machines/labeled transition systems distinguish themselves from these UML diagrams by having the ability to capture the latter, and can even be used to generate sequence diagrams \cite{5773385}.

Capturing the exact behavior of any given program is impossible as it simply reduces to the halting problem.
As such, behavioral extraction techniques that remain generally applicable either have to make strong assumptions about the underlying software or approximate the behavior of the system.

In particular, so-called static extraction techniques focus on generating models that over-approximate the system behavior without running the underlying software.
In such cases, the effect that data has on the control flow of the component is approximated using a subset of semantics rules.
For example, in \cite{kollmann2001capturing,10.1145/1062455.1062510,1235418}, the respective authors introduce techniques for the automatic extraction of UML interaction and/or sequence diagrams from code using (partial) flow analysis to approximate the behavior of the system for any possible input.
While these models capture the behavior of individual function calls, the extracted models are effectively stateless and thus do not allow us to distinguish different states within the extracted component.
This is caused by the techniques not being able to capture the precise data flow of an actual execution and thus the actual control flow, which is influenced by said data, must be approximated, e.g.\ through symbolic executions. 

Dynamic extraction techniques, which are the dual counterpart of static techniques, examine ongoing or previous executions of the program to construct a model.
Using actual executions, these techniques can precisely capture data as it flows through the program.
However, this means that it can only capture the behavior that is observed during the execution.
As such, the model generated is an under-approximation of the system behavior and its completeness is severely dependent on the inputs being supplied, and edge cases might not become visible within the model, e.g.\ data races, or issues relating to timing.

Regardless of their potential incompleteness, dynamic extraction techniques have repeatedly shown to be useful for understanding complex software systems.
For example, in \cite{systa2001shimba}, Systä et al.\ introduce the Symbo environment, which uses a modified Java SDK to capture execution traces, which are used to generate both UML sequence and state diagrams.

An additional benefit of dynamic techniques is that no access to the source code is needed, e.g.\ the reverse engineering of legacy systems.
The key challenge here however is constructing a compact finite state machine allowing for traces of arbitrary length, whilst only having access to finite traces.
Various variations of the k-tail algorithm \cite{5009015}, such as the approaches in \cite{5773385,7728088}, or the MINT tool \cite{walkinshaw2016inferring}, which is based on Evidence-Driven State Merging \cite{lang1998results}, have shown to be effective in this regard.

Active learning techniques, originally introduced by Angluin in \cite{angluin1987learning}, are another example of dynamic extraction techniques in which no access to the source code is needed.
In this, a state machine, the so-called hypothesis model, is repeatedly constructed based on inputs and outputs sent to and from the software component.
Whenever a trace of inputs and outputs results in a counterexample that does not conform to the hypothesis model, the model is updated to include the discrepancy.
This is repeated till no more counterexamples can be found.

A recent example of these techniques being used for complex software systems is \cite{aslam2020interface}, in which state-based interfaces are extracted by learning Java and C++ code and combined with logs of previous executions.
Learning is also commonly used within cyber security to infer state machines of communication protocols  \cite{de2015protocol,comparetti2009prospex}, often referred to as fuzzing.
However learning models from code is computationally expensive, as the amount of tests required to test for counterexamples increases exponentially by the number of states of the model \cite{vaandrager2017model} and can thus only be applied to systems with smaller models, e.g.\ the interfaces extracted in \cite{aslam2020interface} all contain less than a $1000$ states.
As such, learning often requires additional system-specific tricks, abstractions, or rules, to be used effectively on large-scale systems, e.g. in \cite{hooimeijer2022constructive,isberner2014ttt}.
Another example is the so-called grey box fuzzer AFLNet \cite{pham2020aflnet} which combines code-coverage analysis and a pre-known specification model to prioritize tests that most likely lead to counterexamples and thus accelerate the learning of implementation models.

With static and dynamic techniques respectively offering over- and under-approximations, it seems intuitive that the solution lies in a hybrid approach.
In \cite{briand2004towards}, Briand et al.\ draw a similar conclusion and outline a methodology on how static analysis can be used to effectively merge traces obtained through dynamic analysis of the system into UML sequence diagrams for various user stories.
In \cite{hybridApproximationWebApps}, dynamic analysis of the execution of a web application at runtime is combined with the static analysis of the source code of the event handlers found during said interaction, allowing it to capture more applications than other static approaches.

Classifying our approach is perhaps a bit difficult.
The framework itself, i.e.\ the transformation of \cpp code into a process equation that we outline in this paper, is fully static, as no code executions are used to acquire the process equation.
Calculating the associated Labelled Transition System, which is required for its visualization and the application of proving equivalence, is effectively equivalent to actually executing the code.
The main advantage/difference is that all possible executions, within the user-provided bounds, are simulated, thus circumventing the under-approximation issue that dynamic approaches run into.

\section{From Code to Model} \label{sec:framework}
We now discuss our approach to deriving labeled transition systems from \cpp code.
We first give a brief overview of the intermediate language Simple \cpp (SCPP), shown in Figure \ref{fig:scpp_language}.
In Subsection \ref{ssec:process_algebra} we give a concise overview of the process algebra that is used in Subsection \ref{ssec:semantics} to formalize the semantics of SCPP, i.e.\ the process algebraic equations that are used to generate a labeled transition system from any SCPP fragment.
Last, in Subsection \ref{ssec:transformation} we discuss the transformation of \cpp code into this SCPP language.

\begin{figure}
    \centering
    $\begin{array}{l}
    \begin{array}{l c l}
    \textit{Scpp} & := & \textbf{return}(\textit{E}) \\
        & | & \textbf{assign}(\textit{Scope}, \textit{ID}_x, \textit{E}) \\
        & | & \textbf{call}(\textit{ID}_x, E, \textit{ID}_f, \textit{List}(E),\\
        &   & \quad \textit{List}(\textit{ID}_x)) \\
        & | & \textbf{ite}(E, \textit{List}(\textit{Scpp}), \textit{List}(\textit{Scpp})) \\
        & | & \textbf{while}(E, \textit{List}(\textit{Scpp})) \\
%        & | & \textbf{for\_all}(\textit{ID}, E) \\
        & | & \textbf{ref\_load}(\textit{ID}_x, E) \\
        & | & \textbf{ref\_assign}(\textit{ID}_x, \textit{E}) \\
        & | & \textbf{jump}(\textit{Flag}, \textit{Lbl}) \\
        & | & \textbf{flag}(\textit{Flag}, \textit{Lbl}) \\
        & | & \textbf{throw} \\
        & | & \textbf{catch}(\textit{List}(\textit{Scpp})) \\
        & & \quad \textit{List}(\textit{ID}_x), \textit{List}(\textit{Scpp})) \\
        & | & \textbf{call\_lambda}(\textit{ID}_x, \textit{ID}_x, \textit{List}(E),\\
        &   & \quad \textit{List}(\textit{ID}_x)).
    \end{array} \\ 
    \\ 
    \textit{Scope} := \textbf{local} | \textbf{global} \\
    \textit{Flag} := \textbf{continue} | \textbf{break}\\
    \end{array}
    \quad 
    \begin{array}{l}
    \begin{array}{l c l}
    \textit{E} & := & \textbf{minus}(E, E) \\
        & | & \textbf{plus}(E, E) \\ 
        & | & \textbf{divide}(E, E) \\
        & | & \textbf{multiply}(E, E) \\
        & | & \textbf{equals}(E, E) \\
        & | & \textbf{not\_equals}(E, E)\\
        & | & \textbf{greater\_than}(E, E)\\
        & | & \textbf{greater\_equal}(E, E)\\
        & | & \textbf{smaller\_equal}(E, E)\\
        & | & \textbf{smaller\_than}(E, E) \\
        & | & \textbf{not}(E) \\
        & | & \textbf{or}(E, E) \\
        & | & \textbf{and}(E, E) \\
        & | & \textbf{read}(\textit{Scope}, \textit{ID}_x) \\
        & | & \textbf{constant}(V) \\
        & | & \textbf{at}(E, E) \\
        & | & \textbf{self} \\
        & | & \textbf{ref\_field}(E, \textit{ID}_x) \\
        & | & \textbf{init\_lambda}(\textit{List}(\textit{ID}_x),\\
        &   & \quad \textit{List}(\textit{ID}_x), \textit{List}(\textit{ID}_x),\\
        &   & \quad \textit{List}(\textit{Scpp})) \\
        & | & \textbf{init\_list}(\textit{List}(E)).
    \end{array} \\ \\ \begin{array}{l c l}
    \textit{V} & := & \textbf{Number}(\mathbb{Z}) \\
        & | & \textbf{Boolean}(\mathbb{B}) \\
        & | & \textbf{OrderedSet}(\textit{List}(V)) \\
        & | & \textbf{VoidType} \\
        & | & \textbf{PType}(\textit{ID}_p) \\
        & | & \textbf{EnumType}(\textit{Enum}) \\
        & | & \textbf{FieldRef}(\textit{ID}_p, \textit{ID}_x) \\
        & | & \textbf{LocalRef}(\textit{ID}_x) \\
        & | & \textbf{LambdaType}(\textit{List}(\textit{ID}_x),\\
        &   & \quad \textit{List}(\textit{ID}_x \times V), \textit{List}(\textit{ID}_x), \\
        &   & \quad \textit{List}(\textit{Scpp})). \\
        & | & \textbf{StringType}(\textit{List}(\textit{StrLtrl})) 
    \end{array} \end{array}$
    \caption{The \textit{SCPP} language and associated sub-languages and data-types.}
    \label{fig:scpp_language}
\end{figure}

At the core of the SCPP language, shown in Figure \ref{fig:scpp_language}, we have three sets: the set of statements $\textit{Scpp}$, the set of expressions $E$, and the set of values $V$.
We explain the meaning of each statement and expression in Subsection \ref{ssec:semantics} when discussing the semantics.
In general, the idea is that $\cpp$ statements are mapped to $\textit{Scpp}$, and expressions to the set $E$.
Constants and values calculated during execution are stored in $V$.
The precise distinction, however, lies in the semantics of the two; that being, the evaluation of an \textit{Scpp} statement leads to one or more transitions being taken and changes to our state, whereas the evaluation of an expression yields only a value, described by the set $V$, and does not introduce any side-effects.

Besides these two sets, we make use of additional data types.
The set of identifiers $\textit{ID}$ is a simple enumerable set of identifiers. 
We often annotate the set and its elements with an additional subscript to clarify typing, i.e.\ $\textit{id}_x$ for variables, $\textit{id}_p$ for processes, and $\textit{id}_f$ for function names.
The elements in the set $\textit{Scope}$ are used to distinguish variables.
The \textbf{global} scope distinguishes variables, i.e.\ fields, remain relevant even after a function terminates.
The \textbf{local} scope distinguishes local variables, i.e.\ method parameters or variables declared within a method body, which lose relevance upon termination or during a nested function call.
The sets $\textit{Flag}$, and \textit{Lbl} are used for handling preemptive breaking of loops using the \textbf{continue} and \textbf{break} statements, and are further explained in Subsection \ref{ssec:transformation}.
The set $\textit{Enum}$ contains any possible \textit{Enum} used by the program and is populated during transformation with the enums defined within the transformed translation unit.
Besides this we us a \textit{List} data type, with the classical head ($\textit{hd}$), tail ($\textit{tl}$), cons ($::$), and concatenate ($\cat$) operators. 
The empty list is denoted as $\emptylist$.

\subsection{Introduction to Process Algebra} \label{ssec:process_algebra}
We specify the precise formal semantics of SCPP using a process equation similar to those in mCRL2 \cite{mcrl2}, which is based on the Algebra of Communicating Processes \cite{DBLP:books/daglib/0069083}, and Calculus of Communicating Processes \cite{DBLP:books/sp/Milner80}.
Generally, process equations come in the form $P(d:D) = \hdots~$, where the left hand side of an equation is a process variable over some set of states $D$, specifying, on the right hand side, what actions can occur given that we are in some state $d\in D$, and what the resulting states are.
Actions represent any sort of atomic event such as calling a function, or writing to a field.
An action consists of a label and a possible set of data parameters, e.g.\ the action $\texttt{call}(\textit{actuator}, \textit{move}, 1)$ has $\textit{call}$ as the label and $\textit{actuator}, \textit{move}$, and $1$ as data parameters.
Process states and data parameters can be of varying types such as booleans, algebraic data types, such as the ones defined in Figure \ref{fig:scpp_language}, and mappings.
A special action $\tau$, the so-called hidden or internal action, is used to represent an action that is externally not directly visible.

Process expressions, i.e.\ the right hand side of the process equation, follow the following syntax:
$$p ::= \alpha ~|~ X(e) ~|~ p \cdot p ~|~ p + p ~|~ (c) \to p \diamond p ~|~ \sum_{d{:}D} p ~|~ \textbf{match}~e~\textbf{with}~(d \to p)^+$$
Here, $\alpha$ is an action, and $X(e)$ denotes recursion to some process variable $X$ with the evaluation of data expression $e$ as its new state.
We use $p\cdot p$ to model sequential composition, e.g.\  $P(b{:}\mathbb{B}) = o(b) \cdot P(\neg b)$ models a process that interleaves the actions $o(\textit{true})$ and $o(\textit{false})$. 
Non-deterministic choice is modelled through the (+) operator.
Whereas, $\sum_{d{:}D}$ models the non-deterministic choice of picking any data element $d$ of type $D$.
The expression $(c) \to p \diamond p'$ models the classic if-then-else, by stating that $p$ can occur iff the boolean expression $c$ evaluates to true, and $p'$ can occur iff $c$ evaluates to false.

Additionally, towards improving the readability of our process equations, we introduce the expression $\textbf{match}~e~\textbf{with}~(d \to p)^+$.
This expression models the evaluation of the expression e and matches it with some data expression $d$.
We then proceed with the process expression $p$ associated with said data expression $d$.
We note that this expression can also be written using a combination of the previous process expressions.

\subsection{From Data Structure to Model} \label{ssec:semantics}
\begin{figure}
    \centering
    \begin{tikzpicture}
    \node[draw, fill=lightgray, minimum height=1.3cm, minimum width = 2.3cm] (Stable) {Stable ($\Sigma$)};
    \node[draw, below = 2 of Stable, align=center, minimum height=1.3cm, minimum width = 2.3cm] (Processing) {Processing \\(State)};

    \draw[<-] (Stable.north) -- ++(0,.3cm); 
    
    \draw[->] ([yshift=.5cm]Stable.east) -- ++(.8cm,0) -- node[midway, right, align=left] {
        load($\hdots$) \\
        store($\hdots$)
    } ++(0,-1cm) -- ++(-.8cm, 0);

    \draw[->] ([yshift=.5cm]Processing.east) -- ++(.8cm,0) -- node[midway, right, align=left] {
        load($\hdots$) \\
        store($\hdots$) \\
        call\_func($\hdots$)
    } ++(0,-1cm) -- ++(-.8cm, 0);

    \draw[->] ([xshift=.8cm]Stable.south)-- node[midway, right, align=left] {
        call\_func($\hdots$)
    } ([xshift=.8cm]Processing.north);
    \draw[<-] ([xshift=-.8cm]Stable.south)-- node[midway, left, align=right] {
        return\_func($\hdots$) \\
        throw\_func($\hdots$)
    } ([xshift=-.8cm]Processing.north);
    
    \end{tikzpicture}
    \caption{This diagram presents an abstract overview of the processes described in this section.}
    \label{fig:example}
\end{figure}
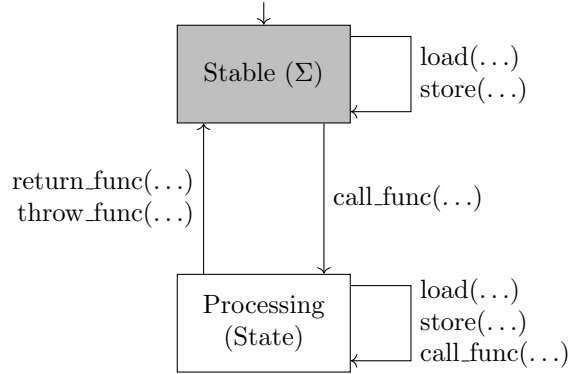

Before we discuss the precise formal semantics of SCPP, let us reconsider the example in Figure \ref{fig:example}.
When we formalize the behavior of a given class instance under run-to-completion semantics, we can distinguish two distinct sets of states: stable and processing states, which are distinguished by whether or not a function is being executed.
Processes start out in a stable state in which their fields can be read and written to, but no internal calculation takes place.
Once one of its functions is called, signified by the call\_func action, the function body is executed, resulting in potential loads, stores, and function calls to its members. 
When it eventually terminates through a return or if an exception is thrown, it returns to a (potentially new) stable state.

The set of stable states, i.e.\ no function is being executed, is defined as: $\Sigma = \textit{ID}_p \times V^{\textit{ID}_x}$, where $\textit{ID}_p$ is the set of process identifiers used to distinguish different object instances, and $V^{\textit{ID}_x}$ maps a given instances fields to their current value. 
The set of processing states, $\textit{State} = \textit{ID}_p \times  V^{\textit{ID}_x} \times \textit{Stack}$, is more complex as we need it to support the various features of modern-day programming languages, such as recursion and pass-by-reference, while also needing to keep track of the \textit{Scpp} instructions that need to be executed.
Towards this, we define the call stack data structure (\textit{Stack}) in Definition \ref{def:stack}, to keep track of both the current to-be-executed instructions and any to-be executed instructions of previous stack frames in case of nested function calls.

\begin{definition} \label{def:stack}
We define the set of call stacks $\textit{Stack}$ as the smallest set such that:
\begin{itemize}
    \item Given any function name $f_{\textit{id}} \in \textit{ID}$ we have $\textbf{external}~f_{\textit{id}} \in \textit{Stack}$,
    \item given any program $\textit{prog}\in \textit{List(S)}$, variable name $x \in \ID$, state mapping $\stloc \in \textit{State}$, local references $\textit{refs} \in \textit{List}(\textit{ID}_x \times V)$, and call stack $\textit{stl} \in \textit{Stack}$ we have $\textbf{frame}(\textit{prog}, x, \stloc, \textit{refs}) :: \textit{stl} \in \textit{Stack}$.
\end{itemize}
\end{definition}
\noindent The variable \textit{prog} is a list of to-be-executed instructions, $x$ is a variable to which the result of the current method should be written, $\stloc$ maps each \textbf{local} variable to its current value, $\textit{refs}$, which we will explain in further detail later, is used to keep track of any passed-by-reference variables, and $\textit{stl}$ is the tail of the current call stack.
Elements of the form $\textbf{external}~f_\textit{id}$ indicate that the call stack, which belongs to calling the method $f_\textit{id}$, has terminated. 

Whenever a nested function call occurs, we must keep track of any local variables (and their assigned values) which are passed-by-reference.
This is done through so-called local-reference-assignments (LRAs), which are simple tuples consisting of the passed-by-reference variable and their current value.
In Definition \ref{def:LRA_mappings}, we define some mappings that we use to easily work with the list of LRAs.
When we enter a nested function call, the \textit{makeLRAs} mapping is used to construct the list of LRAs, using their current values.
Reading the values of these variables, is done through the $\textit{readLRA}$ mapping, returning the value of the matching identifier.
Updating the values of passed-by-reference variables, is done through the $\textit{updateLRA}$ mapping.
When the nested function call terminates, the previous local variables are updated by updating it with the tracked LRAs using $\textit{ret\_refs}_\sigma$. 
In the case that the previous variable, is also holding a reference, the LRA is propagated through the $\textit{ret\_refs}_r$ mapping.

\begin{definition} \label{def:LRA_mappings}
    We define the following set of mappings for LRAs:
    \begin{itemize}
    \item $\textit{makeLRAs} : \textit{List}(\textit{ID}_x) \times (\State) \to \textit{List}(\textit{ID}_x \times V)$ such that\\ 
        $\textit{makeLRAs}([\textit{id}_1, \hdots, \textit{id}_n], \sigma) = [\langle \textit{id}_1, \sigma(\textit{id}_1) \rangle, \hdots, \langle \textit{id}_n, \sigma(\textit{id}_n) \rangle]$,

    \item $\textit{readLRA} : \textit{ID}_x \times \textit{List}(\textit{ID}_x \times V) \to V $ such that \\
        $\textit{readLRA}(\textit{id}_x, \langle \textit{id}_x', v\rangle :: \textit{lras}) 
        = \left\{\begin{array}{l l}
            v & \textbf{if}~(\textit{id}_x = \textit{id}_x') \\
            \textit{readLRA}(\textit{id}_x, \textit{lras}) & \textbf{otherwise} , 
        \end{array} \right. $
        
    \item $\textit{updateLRA} : \textit{ID}_x \times V \times \textit{List}(\textit{ID}_x \times V) \to \textit{List}(\textit{ID}_x \times V)$ such that \\
        $\begin{array}{l c l}
        \textit{updateLRA}(\textit{id}_x, v, []) & = & [], \\
        \textit{updateLRA}(\textit{id}_x, v, \langle \textit{id}_x', v' \rangle :: \textit{lras}) & = & \langle \textit{id}_x', v \rangle :: \textit{lras} ~ \text{iff}~(\textit{id}_x \approx \textit{id}_x'),\\
        \textit{updateLRA}(\textit{id}_x, v, \langle \textit{id}_x', v' \rangle :: \textit{lras}) & = & \langle \textit{id}_x', v' \rangle :: \textit{updateLRA}(\textit{id}_x, v, \textit{lras}) ~ \textbf{otherwise},\\
        \end{array}$

    \item $\textit{ref\_refs}_\sigma : \textit{List}(\textit{ID}_x \times V) \times (\State) \to (\State)$ such that \\
    $\begin{array}{l l}
        \textit{ret\_refs}_\sigma([], \sigma) &= \sigma\\
        \textit{ret\_refs}_\sigma(\langle \textit{id}, v \rangle :: \textit{lras}, \sigma) &= \left\{\begin{array}{l l}
            \textit{ret\_refs}_\sigma(\textit{lras}, \sigma) 
            &    \textbf{if}~\sigma(\textit{id}) = \textbf{LocalRef}(\textit{id'}) \\
            \textit{ret\_refs}_\sigma(\textit{lras}, \sigma)
            &    \textbf{if}~\sigma(\textit{id}) = \textbf{FieldRef}(\textit{id}_p, \textit{id}_x) \\
            \textit{ret\_refs}_\sigma(\textit{lras}, \sigma[\textit{id} \mapsto v])
            &   \textbf{otherwise} \text{, and}
        \end{array} \right.
    \end{array}$

    \item $\textit{ret\_refs}_r : \textit{List}(\textit{ID}_x \times V) \times (\State) \to \textit{List}(\textit{ID}_x \times V)$ such that \\
    $\begin{array}{l l}
        \textit{ret\_refs}_r([], \sigma, \textit{lras}) &= \textit{lras} \\
        \textit{ret\_refs}_r(\langle \textit{id}, v \rangle :: \textit{lras}', \sigma, \textit{lras}) &=  
        \left\{\begin{array}{l l}
            \textit{updateLRA}(\textit{id}', v, & \\
            \multicolumn{1}{r}{\quad \quad \quad \quad \quad \textit{ret\_refs}_r(\textit{lras}', \sigma, \textit{lras}))}
            &   \textbf{if}~\sigma(\textit{id}) = \textbf{LocalRef}(\textit{id}') \\
            \textit{ret\_refs}_r(\textit{lras}', \sigma, \textit{lras})
            &   \textbf{otherwise}.
        \end{array}\right.\\
    \end{array}$
    \end{itemize}
\end{definition}

As we have two distinct sets of states, so do we have two distinct process equations.
One process equation for stable states, defined in Definition \ref{def:semantics_stable}, and one process equation for processing, outlined in Definition \ref{def:semantics_Scpp} in Appendix \ref{sec:appendixA}.
The semantics for evaluating expressions in $E$ is outlined in Definition \ref{def:semantics_Eval}. 
Both process equations use only a limited set of action labels, given in Definition \ref{def:action_labels}.

The actions defined in Definition \ref{def:action_labels}, with the exception of the $\tau$ action, consist of two parts: namely a top, and a bottom part indicated by $\texttt{\_t}$ and $\texttt{\_b}$ suffixes respectively.
Through the usage of the so called communication and allow operators of our process algebra, we can enforce communication between these two parts.
That is, a top action and bottom action can only occur if the other half also occurs simultaneously, with the exact same data parameters, e.g.\ the $\texttt{store\_b}(p_{\textit{to}}, \textit{field}, \textit{newValue})$ can only occur in the process $C$, if another process, e.g.\ an instance of process $P$, takes the $\texttt{store\_t}(p_{\textit{to}}, \textit{field}, \textit{newValue})$ action.
When these two actions occur simultaneously, we instead write the third action label, e.g.\ $\texttt{store\_comm}$.

\begin{definition} \label{def:action_labels}
    We define the following action labels:
    \begin{itemize}
        \item $\texttt{call\_func\_t}, \texttt{call\_func\_b}, \texttt{call\_func} : \textit{ID}_p \times \textit{ID}_f \times \textit{List}(V) \times \textit{List}(\textit{ID}_x \times V)$ representing function calls,
        \item $\texttt{return\_func\_t}, \texttt{return\_func\_b}, \texttt{return\_func} : \textit{ID}_p \times \textit{ID}_f \times V \times \textit{List}(\textit{ID}_x \times V)$ representing successful function termination,
        \item $\texttt{throw\_func\_t}, \texttt{throw\_func\_b}, \texttt{throw\_func} : \textit{ID}_p \times \textit{ID}_f \times \textit{List}(\textit{ID}_x \times V)$ representing a function throwing an error,
        \item $\texttt{load\_t}, \texttt{load\_b}, \texttt{load\_comm} : \textit{ID}_p \times \textit{ID}_x \times V$ representing reading the value of a given instances field,
        \item $\texttt{store\_t}, \texttt{store\_b}, \texttt{store\_comm} : \textit{ID}_p \times \textit{ID}_x \times V$ representing writing a value being written to a given instances field, and
        \item the $\tau$ action which abstracts over any sort of internal calculation.
    \end{itemize}
\end{definition}

\begin{definition}\label{def:semantics_stable}
We define the process equation for stable states $C$ as follows:

$\begin{array}{l}
 C(\langle \pid, \stglob \rangle : \Sigma) = \\
 \quad \sum_{f{:}\textit{ID}_f, \textit{vs}{:}\textit{List}(V), \textit{lras}{:}\textit{List}(\textit{ID}_x \times V)} .\texttt{call\_func\_b}( \pid, f, \textit{vs}, \textit{lras}) \cdot \\
 \quad \quad P(\pid, \stglob, \STF{\textit{get\_prog}(f)}{\bot}{i_{\stloc}(f, \textit{vs})}{\textit{lras}}{\textbf{external}~f}) \\
 \quad + \sum_{\textit{id}{:}\textit{ID}_x} . \texttt{load\_b}(\pid, \textit{id}, \stglob(\textit{id})) \cdot C(\pid, \stglob) \\
 \quad + \sum_{\textit{id}{:}\textit{ID}_x, v{:}V}. \texttt{store\_b}(\pid, \textit{id}, v) \cdot C(\pid, \stglob[ \textit{id} \mapsto v])
\end{array}$ \\
where
\begin{itemize}
    \item $\textit{get\_prog} : \textit{ID}_f \to \textit{List}(\textit{Scpp})$ maps every function id to the list of SCPP instructions that have been acquired by transforming its body using the transformation mappings in the next subsection, and
    \item $i_{\stloc} : \textit{ID}_f \times \textit{List}(V) \to (\textit{ID}_x \to V)$ gives the initial local variable assignment in which the list of arguments are assigned to the associated variables of the given function. 
    \end{itemize}
\end{definition}

The stable process $C$ accepts three types of operations: function calls, and reading and writing to the fields of the class instance.
When a function call is accepted, we proceed to the processing process $P$ with the list of SCPP instructions matching the called function and the initial local variable assignment matching the arguments.
Field accesses simply either read or update the field mapping of our process.

The evaluation of expressions is done using the mapping defined in Definition \ref{def:semantics_Eval}.
We do not list all binary operators, as these all follow the same structure.
With the exception of the \textbf{ref\_global}, and \textbf{init\_lambda} expressions, the evaluation of most expressions is straightforward.
The $\textbf{ref\_global}(e_1, \textit{id}_x)$ expression is a field reference expression, where $e_1$ is the expression evaluating to the process owning the field $\textit{id}_x$.
The $\textbf{init\_lambda}(\textit{ids}, \textit{copies}, \textit{refs}, S)$ initializes a lambda function, with arguments $\textit{ids}$, captures-by-copy \textit{copies}, captures-by-reference \textit{refs}, and function body \textit{S}.
Calling such a lambda function is explained later.

\begin{definition} \label{def:semantics_Eval}
We define the expression evaluation semantics as the following mapping $\semE : E \times \textit{State} \to V$.
Let $st = \langle \pid, \stglob, \STF{\textit{prog}}{x}{\stloc}{\textit{refs}}{\textit{stl}} \rangle$ in \\
    
    $\begin{array}{l l}

    \semE(\textbf{plus}(e_1, e_2), \st) & = \textbf{Number}(n_1 + n_2)\\
        & \begin{array}{l l l}
            \whr & \textbf{Number}(n_1) &= \semE(e_1, \st) \\
             & \textbf{Number}(n_2) &= \semE(e_2, \st)
        \end{array} \\
    
    \semE(\textbf{equals}(e_1, e_2), \st) &= \textbf{Boolean}(\semE(e_1, \st) = \semE(e_2, \st) ) \\
        
    \semE(\textbf{not}(e)) &=  \textbf{Boolean}(!b) \\
        & ~~ \whr \textbf{Boolean}(b) = \semE(e, \st) \\

    \semE(\textbf{read}(\textbf{Local}, \textit{id}_x), \st) &= \stloc(\textit{id}_x) \\
    \semE(\textbf{read}(\textbf{global}, \textit{id}_x), \st) &= \stglob(\textit{id}_x) \\

    \semE(\textbf{constant}(v), \st) &= v \\

    \semE(\textbf{at}(e_1, e_2), \st) &= \textit{vs} . n \\
        & \begin{array}{l l l}
            \whr & \textbf{OrderedSet}(\textit{vs}) & = \semE(e_1, \st)\\
             & \textbf{Number}(n) & = \semE(e_2, \st)
        \end{array}\\

    \semE(\textbf{self}, \st) &= \textbf{PType}(\pid) \\

    \semE(\textbf{ref\_global}(e_1, \textit{id}_x), \st) &= \textbf{FieldRef}(\textit{id}_p', \textit{id}_x) \\
        & ~~ \whr \textbf{PType}(\textit{id}_p') = \semE(e_1, \st) \\

    \semE(\textbf{init\_lambda}(\textit{ids}, \textit{copies}, \textit{refs}, S), \st) &= \textbf{LambdaType}(\textit{ids}, \textit{lras}, \textit{refs}, S)\\
        & ~~ \whr \textit{lras} = \textit{make\_LRAs}(\textit{copies}, \stloc) \\

    \semE(\textbf{init\_list}(\textit{es})) &= \semE^*(\textit{es}, \st)\\
    \end{array}$\\
    \text{where} $\semE^*([e_1, \hdots e_n], \st) = [\semE(e_1, \st), \hdots, \semE(e_n, \st)]$
    
\end{definition}

Let us now discuss the process equation $P$, and the \textit{Scpp} instructions it processes, in further detail.
Since the process equation is quite large, we split up the process equation, and discuss each part.
The process equation is shown in its entirety in the Appendix.

The process $P$ repeatedly takes the head of our list of \textit{Scpp} instructions and acts accordingly, till either the list of instructions is empty or a $\textbf{return}$ instruction is encountered.
When either happens, the helper process $P_{\textit{return}}$ takes the to-be returned value $v$ and inspects the current tail of the call stack.
If there is a previous stack frame, said stackframe becomes the current stackframe, if there are no further instructions, a $\texttt{return\_func\_b}$ action occurs and we return to the stable process $C$.
In the case of a $\textbf{return}(e)$ statement, the expression $e$ is first evaluated and passed on either as the return value or to the associated variable in the previous stack frame.

$$
\begin{array}{l}
% begin P process equation
P(\st = \langle \pid : \ID, \stglob : \State, \textit{call\_stack} {:} \textit{List(StackFrame)} \rangle) = \textbf{match}~\textit{call\_stack}~\textbf{with} \\
    \tab{}{1} \STF{\textit{prog} : \textit{List(S)}}{x : \textit{ID}}{\stloc}{\textit{refs} : \textit{List}(\textit{ID} \times V)}{stl} \to (\textit{prog} \approx [])\\
    \tab{}{2} \to P_{\textit{return}}(\textbf{VoidType}, \st) \\
    \tab{}{2} \diamond ~\textbf{match} ~ \textit{hd(prog)} ~ \textbf{with}\\
    % begin individual scpp statement semantics
    \tab{}{3}\begin{array}{l c l}
    \textbf{return}(e) & \to & P_{\textit{return}}(\semE(e,\st), \st) \\
    \quad \vdots
    \end{array} \\
\text{where} \\
P_{\textit{return}}(v:V, \st = \langle \pid : \ID, \stglob : \State, \textit{call\_stack} {:} \textit{List(StackFrame)} \rangle) = \textbf{match}~\textit{stl}~\textbf{with}\\
    \quad \begin{array}{l c l}
        \textbf{external}~f_{\textit{id}} & \to & \texttt{return\_func\_b}(f_{\textit{id}}, v, \textit{refs}) \cdot C(\pid, \stglob)  \\
        \STF{\textit{prog}'}{x'}{\stloc'}{\textit{refs}'}{stl'} & \to & \tau {\cdot} P(\pid, \stglob, \STF{\textit{prog}'}{x'}{\textit{ret\_refs}(\textit{refs}, \stloc'[x \mapsto v])}{\textit{refs}'}{stl'}) 
    \end{array}
\end{array}
$$

We now discuss the $\textbf{match} ~ \textit{hd(prog)} ~ \textbf{with}$ cases for the other \textit{Scpp} instructions.
As such, from here on out, we assume that $\textit{call\_stack} = \STF{\textit{prog}}{x}{\stloc}{\textit{refs}}{\textit{stl}}$ and that $\textit{prog}$ is non-empty.
Let us also define the following shorthand $\textit{nextframe} = \STF{\tlprog}{x}{\stloc}{\textit{refs}}{\textit{stl}}$.

Assigning the evaluation of some expression $e$ to a non-reference variable $y$ is done using the instruction $\textbf{assign}(\textit{scope}, y, e)$.
The $\textit{scope}$ variable, as discussed earlier, is used to distinguish between assignments to local variables and a processes own field.
When an $\textbf{assign}$ instruction is encountered, the expression is evaluated and assigned to the mapping of the associated scope.

The standard control-flow instructions \textbf{ite} (if-then-else) and \textbf{while} instructions are rather straightforward, consisting of some condition $c$ that is evaluated, and a a list of instructions that are added to our list of instructions according to the evaluation of the condition.
For the \textbf{ite} instruction, either the then-body $\textit{prog}_T$ or else-body $\textit{prog}_F$ are added to the front of the list of instructions.
For the \textbf{while} instruction, given that the condition evaluates to true, the while-body $\textit{prog}_W$ is added to to the front of the list of instructions. However, the \textbf{while} instruction itself is not removed, to allow for possible future iterations of the same loop.

$$
\begin{array}{l}
\tab{}{5} \vdots \\
\quad \begin{array}{l c l}
    
    \textbf{assign}(\textbf{local}, y, e) & \to &  \tau \cdot P(\pid, \stglob, \STF{\tlprog}{x}{\stloc[y \mapsto \semE(e, \st)]}{\textit{refs}}{stl}) \\
    
    \textbf{assign}(\textbf{global}, y, e) & \to & \tau \cdot P(\pid, \stglob[y \mapsto \semE(e, \st)], \STF{\tlprog}{x}{\stloc}{\textit{refs}}{stl}) \\
    
    \textbf{ite}(c, \textit{prog}_T, \textit{prog}_F) & \to & (\semE(c, \st)) \\
    & & \quad \begin{array}{c l}
        \to & \tau \cdot P(\pid, \stglob, \STF{\textit{prog}_T \cat \tlprog}{x}{\stloc}{\textit{refs}}{stl})\\
        \diamond & \tau \cdot P(\pid, \stglob, \STF{\textit{prog}_F \cat \tlprog}{x}{\stloc}{\textit{refs}}{stl})\\
        \end{array}\\

    \textbf{while}(c, \textit{prog}_W) & \to & (\semE(c, \st)) \\
    & & \quad \begin{array}{c l}
        \to & \tau \cdot P(\pid, \stglob, \STF{\textit{prog}_W \cat \textit{prog}}{x}{\stloc}{\textit{refs}}{stl})\\
        \diamond & \tau \cdot P(\pid, \stglob, \textit{nextFrame})\\
        \end{array}\\

\end{array} \\
\tab{}{5} \vdots \\
\end{array}
$$

The function call instruction $\textbf{call}(y, e_p, \fid, \textit{es}, \textit{refs})$ states that function $\fid$ belonging to the evaluation of $e_p$ should be called using the evaluation of $\textit{es}$ as arguments and $\textit{refs}$ as passed references, and the result of this function call should be assigned to the local variable $y$.
Thus, the function owner expression $e_p$ is first evaluated; if the result matches our own process identified $\pid$, no externally observable behavior will be able to be observed and a new stack frame, matching the called function, is added to our call stack.
If the result of the expression is of a different process identifier, i.e. $\pid'$, the $\texttt{call\_func\_t}$ action occurs, which enforces communication with the other process with the process identifier $\pid'$.
Execution of the called function is then handled by said process, and, unless no termination occurs, will eventually lead to either a $\texttt{return\_func\_b}$ or $\texttt{throw\_func\_b}$ action.
In the case of successful termination, our local state is updated using the result(s) of the function call.
$$
\begin{array}{l}
\tab{}{5}\vdots \\
\tab{}{1}\textbf{call}(y, e_p, \fid, \textit{es}, \textit{refs}) \to \textbf{match}~\semE(e_p, \st)~\textbf{with}\\
\tab{}{2} \begin{array}{l l l @{} l}
            \textbf{PType}(\pid) & \to &
            \tau \cdot P(\pid, \stglob, \textbf{frame}( & \textit{get\_prog}(\fid), y, \\
            & & & i_{\stloc} (\fid, \semE^*(\textit{es}, \st)), \\
            & & & \textit{makeLRAs}(\textit{refs}, \stloc)) {::} \textit{nextFrame}) \\
            \textbf{PType}(\pid') & \to & \multicolumn{2}{l}{\texttt{call\_func\_t}(\pid', \fid, \semE^*(\textit{es}, \st), \textit{makeLRAs}(\textit{refs}, \stloc)) \cdot ( }\\
            & & \multicolumn{2}{l}{\quad \sum_{v{:}V, \textit{lras}{:}\textit{List}(\textit{ID}_x \times V)}. \texttt{return\_func\_t}(\pid', \fid, v, \textit{lras}) \cdot}\\
            & & \quad P(\pid, \stglob, \textbf{frame}( & \tlprog, x,\\
            & & & \textit{ref\_refs}_\sigma(\textit{lras}, \stloc[y \mapsto v]),\\
            & & & \textit{ref\_refs}_r(\textit{lras}, \textit{refs})) {::} \textit{stl})\\
            & & \multicolumn{2}{l}{+ \sum_{\textit{lras}:\textit{List}(\textit{ID}_x \times V)}. \texttt{throw\_func\_t}(\pid', \fid, \textit{lras}) \cdot } \\
            & & \multicolumn{2}{l}{\begin{array}{l @{} l}
                P_{\textit{throw}}(\pid, \stglob, \textbf{frame}(& \tlprog, x, \\
                 & \textit{ret\_refs}_\sigma(\textit{lras}, \stloc), \\
                & \textit{refs}) {::} \textit{stl}))
            \end{array}} 
        \end{array} \\
\tab{}{5} \vdots \\
\end{array}
$$

When an error is thrown as a result of a function call or if a \textbf{throw} instruction is processed,  we proceed to the process $P_{\textit{throw}}$.
The $P_{\textit{throw}}$ process simply skips any non-$\textbf{catch}$ statement, til it either finds one or it runs out of instructions to skip.
If a $\textbf{catch}(\textit{prog}')$ instruction is encountered by this process, we first continue with the catch body $\textit{prog}'$ and then the remaining list of (unskipped) instructions.
If no instructions remain, a $\texttt{throw\_func\_b}$ action is taken, allowing the top process to continue with handling the thrown exception, whilst we proceed back to the stable process $C$.
If the $\textbf{catch}$ statement is encountered by the process $P$, it is simply skipped; this corresponds to cases in which a $\textbf{try}~\textbf{catch}$ block is exited succesfully.

$$
\begin{array}{l}
\tab{}{5} \vdots \\
\quad \begin{array}{l c l}
    
    \textbf{throw} & \to & 
        \tau \cdot P_{\textit{throw}}(\st) \\

    \textbf{catch}(\textit{prog'}) & \to &
        \tau \cdot
        P(\pid, \stglob, \textit{nextFrame}) \\

\end{array} \\
\tab{}{5} \vdots \\
\text{where} \\
P_{\textit{throw}}(\st = \langle \pid, \stglob, \textit{call\_stack}\rangle) = \textbf{match}~\textit{call\_stack}~\textbf{with}\\
    \quad \STF{\textit{prog}}{x}{\stloc}{\textit{refs}}{\textit{stl}} \to \textbf{match}~\textit{resolve\_throw}(\textit{prog})~\textbf{with}\\
    \quad \quad \begin{array}{l c l}
        \textbf{framebreak} & \to & 
        \textbf{match}~\textit{stl}~\textbf{with}\\
        \multicolumn{3}{l}{\quad\begin{array}{l c l}
            \textbf{external}~\fid & \to &
            \texttt{throw\_func\_b}(\pid, \fid, \textit{refs}) \cdot C(\pid, \stglob)\\
            \STF{\textit{prog}'}{x'}{\stloc'}{\textit{refs}'}{\textit{stl}'} & \to &
            \tau \cdot P_{\textit{throw}}(\pid, \stglob, \STF{\textit{prog}'}{x'}{\stloc'}{\textit{refs}'}{\textit{stl}'})
         \end{array}} \\
        
        \textbf{caught}(\textit{prog}') & \to & 
        \tau \cdot P(\pid, \stglob, \STF{\textit{prog}' \cat \tlprog}{x}{\stloc}{\textit{refs}}{\textit{stl}})
        \end{array} \\
        
\textit{resolve\_throw}(\emptylist) = \textbf{framebreak} \\
\textit{resolve\_throw}(p {::} \textit{prog}) = \left\{\begin{array}{l l}
    \textbf{caught}(\textit{prog}') & \textbf{if}~p = \textbf{catch}(\textit{prog}')\\
    \textit{resolve\_throw}(\textit{prog}) & \textbf{otherwise}
\end{array}\right.
\end{array}
$$

Reading/writing to a referenced variable is done via the \textbf{ref\_load} and \textbf{ref\_assign} instructions.
The $\textbf{ref\_load}(\textit{id}_x, e_\textit{from})$ instruction states that the expression $e_{from}$ should be dereferenced and stored in the local variable $\textit{id}_x$.
The $\textbf{ref\_assign}(e_\textit{to}, e)$ instruction states that the evaluation of expression $e$ should be stored at the location referenced by $e_\textit{to}$.
The $\textbf{ref\_load}$ and $\textbf{ref\_assign}$ are used to read from/write to a reference.
When processing these instructions, the reference expression, i.e.\ $e_{from}$ or $e_\textit{to}$, is first evaluated to see if a field or a local variable is referenced.
A field reference $\textbf{FieldRef}(\pid', \textit{id}_x')$ points to the field with identifier $\textit{id}_x'$ of the process $\pid'$, and is accessed by communicating using a $\texttt{load\_t}$, or $\texttt{store\_t}$ action.
A local variable reference $\textbf{LocalRef}(\textit{id}_x')$ points to the LRA $\textit{id}_x'$ in our current stack frame, and is read/written to using one of the mappings defined in Definition \ref{def:LRA_mappings}.

$$
\begin{array}{l}
\tab{}{5} \vdots \\
\tab{}{1}\textbf{ref\_load}(\textit{id}_x, e_\textit{from}) \to \textbf{match}~\semE(e_\textit{from}, \st)~\textbf{with}\\
\tab{}{2} \begin{array}{l c l}
        \textbf{FieldRef}(\pid', \textit{id}_x') & \to & 
        \sum_{v{:}V}. load\_t(\pid', \textit{id}_x', v) \cdot \\
        & & \quad P(\pid, \stglob, \STF{\tlprog}{x}{\stloc[\textit{id}_x \mapsto v]}{\textit{refs}}{stl}) \\
        
        \textbf{LocalRef}(\textit{id}_x') & \to &
        \tau \cdot P(\pid, \stglob, \STF{\textit{tl}(\textit{prog}}{x}{\stloc[\textit{id}_x \mapsto \textit{refE}(\textit{id}_x', \textit{refs})]}{\textit{refs}}{\textit{stl}})
    \end{array}\\

\tab{}{1}\textbf{ref\_assign}(e_{\textit{to}}, e) \to \textbf{match}~\semE(e_{\textit{to}}, \st)~\textbf{with}\\
\tab{}{2}\begin{array}{l c l}
        \textbf{FieldRef}(\pid', \textit{id}_x') & \to & 
        store\_t(\pid', \textit{id}_x', \semE(E', e')) \cdot
        P(\pid, \stglob \textit{nextFrame}) \\
        
        \textbf{LocalRef}(\textit{id}_x') & \to &
        \tau \cdot P(\pid, \stglob, \STF{\tlprog}{x}{\stloc}{\textit{updateLRA}(\textit{id}_x', \semE(e, \st), \textit{refs})}{\textit{stl}}\\ 
    \end{array}\\

\tab{}{5} \vdots
\end{array}
$$

When a $\textbf{jump}(\textit{flag})$ instruction is encountered, we start skipping instructions in our current list of instructions till we encounter a $\textbf{flag}(\textit{flag})$ instruction, with an identical \textit{flag}.
Similarly to the $\textbf{catch}$ intruction, the latter instruction is simply skipped whenever encountered, as its only purpose is for the handling of \textbf{continue}, and \textbf{break} statements through the \textbf{jump} instruction, which we explain in further detail in the next section.

$$
\begin{array}{l}
\tab{}{5} \vdots \\
\quad \begin{array}{l c l}
    
    \textbf{jump}(\textit{flag}) & \to & 
        \tau \cdot
        P(\pid, \stglob, \STF{\textit{jump\_to}(\textit{flag}, \tlprog)}{x}{\stloc}{\textit{refs}}{stl})\\

    \textbf{flag}(\textit{flag}) & \to &
        \tau \cdot
        P(\pid, \stglob, \textit{nextFrame}) \\

\end{array} \\
\tab{}{5} \vdots \\
\text{where} \\
\textit{jump\_to}(\textit{flag}, \textit{prog}) = \left\{ \begin{array}{l l}
    \tlprog & \textbf{if}~\textit{hd}(\textit{prog}) = \textit{flag} \\
    \textit{jump\_to}(\textit{flag}, \tlprog) & \textbf{otherwise}
\end{array}\right.
\end{array}
$$

The $\textbf{call\_lambda}(y, \textit{id}_\lambda, [\textit{arg}_1, \hdots, \textit{arg}_n, \textit{refs})$ instruction is sued for calling lambda functions.
The $\textit{id}_\lambda$ identifier is a local variable that stores the given lambda function, and is first retrieved from the variable mapping.
After this, the process is very similar to calling a recursive/nested function;
the body of the lambda function is put on the stack, the initial variable mapping is initialized using the parameters, i.e.\ $[\textit{par}_1, \hdots, \textit{par}_n]$, and the variables that are captured by copy, i.e.\ $\textit{lras}$, and both the variables that are captured by reference, i.e.\ $[\textit{ref}_1, \hdots, \textit{ref}_m]$, and the variables that are passed by reference, i.e.\ \textit{refs}, are turned into LRAs using the $\textit{makeLRAs}$ mapping,

$$
\begin{array}{l}
\tab{}{5} \vdots \\
\tab{}{1}\textbf{call\_lambda}(y, \textit{id}_\lambda, [\textit{arg}_1, \hdots, \textit{arg}_n], \textit{refs}) \to \textbf{match}~\stloc (\textit{id}_\lambda)~\textbf{with}\\
\tab{}{2} \textbf{LambdaType}([\textit{par}_1, \hdots, \textit{par}_n], \textit{lras}, [\textit{ref}_1, \hdots, \textit{ref}_m], \textit{prog}') \to \\
\tab{}{3}\begin{array}{l l l}
        \tau \cdot P(\pid, \stglob, \textbf{frame}(\!\!\! & \textit{prog}', y & \\
         & \textit{ret\_refs}(\textit{lras}, \stloc \!\!\! & [\textit{par}_1 \mapsto \semE(\textit{arg}_1, \st)] \hdots [\textit{par}_n \mapsto \semE(\textit{arg}_n, \st)] \\
         & & [\textit{ref}_1 \mapsto \textbf{LocalRef}(\textit{ref}_1)] \hdots [\textit{ref}_m \mapsto \textbf{LocalRef}(\textit{ref}_m)]), \\
         & \multicolumn{2}{l}{\textit{makeLRAs} ( [\textit{ref}_1, \hdots, \textit{ref}_m \cat \textit{refs}, \stloc) )) :: \textit{nextFrame}}
    \end{array}
\end{array}
$$

\subsection{From Code to Data Structure} \label{ssec:transformation}
We specify the translation from \cpp to SCPP using three mappings: the mapping $\mathbb{T}_S$ on statements, defined in Definition \ref{def:TransS}, the mapping $\mathbb{T}_E$ on expressions, defined in Definition \ref{def:TransE}, and the mapping $\mathbb{T}_A$ on arguments, defined in Definition \ref{def:TransA}.
These definitions are inductively defined using each other.
We note that arguments require a separate set of transformation rules because variables have distinct semantics when passed-by-reference.

To ensure there is a clear separation between implementation and theory any additional information from the parser that is used to assist in the transformation is abstracted over using the following predicates and mappings:
\begin{itemize}
    \item the predicate $\mathcal{R} : \textit{ID}$, which indicates whether or not a given identifier holds a reference or pointer;
    \item the mapping $\mathcal{S} : \textit{ID}_x \to \textit{Scope}$, mapping each identifier to their respective scope;
    \item the predicate $\varepsilon : \textit{ID}$, which indicates if the given identifier is an enum literal;
    \item the predicate $\mathcal{F} : \textit{ID}$, which indicates if the given identifier is a function;
    \item the predicate $\mathcal{R}_A : \textit{ID}_f \times \mathbb{N}$, which indicates if the $n^{\textit{th}}$ argument of the given function should be passed-by-reference.
\end{itemize}

\noindent In the SSTraGen tool, these predicates and mappings are implemented through querrying the AST and eclipse CDT parser that we use.

We define the statement transformation rules in Definition \ref{def:TransS}, using the mapping $\mathbb{T}_S$.
The secondary argument, mostly denoted as $l$, is used to support jumps, i.e.\ \textbf{continue}, and \textbf{break} statements, and is the label associated with the innermost loop or switch statement at the time.
Whenever a loop or switch statement is transformed, a new label is generated, such that jumps in nested loops and/or switches always jump to the position associated with the innermost loop/switch.
The transformation is then accompanied by \textbf{flag} statements with the new label.
Any \textbf{continue}, or \textbf{break} statement encountered within a loop body is transformed into a \textbf{jump} statement that causes a jump to the associated flag.

The majority of transformation rules for statements simply transform a \cpp statement to either the corresponding \textit{Scpp} statement or rewrite it into a \cpp statement for which we have a straightforward transformation.
The switch statement is perhaps the most intricate one.
When transforming a switch statement, the individual cases and transformed and composed into several \textbf{ite} statements.
Each \textbf{ite} statement contains the body of the current case and all subsequent cases to enable fallthrough.
If a \textbf{break} is encountered inside of a case it will force a jump to the flag at the end of the transformation, thus skipping all subsequent cases.

\begin{definition} \label{def:TransS}
The transformation mapping $\mathbb{T}_S : \textit{Statement} \to \textit{List}(\textit{Scpp})$, which transforms \cpp statements to a list of zero or more Scpp instructions, is defined as follows:
$$\begin{array}{l l}
    \TrS{\{\}} & = \emptylist  \\
    \TrS{\{\textit{St}_1 \hdots \textit{St}_n \} } &= \TrS{\textit{St}_1} \cat \TrS{\{ \textit{St}_2 \hdots \textit{St}_n \}} \\

    \TrS{\textbf{return}~E;} &= \Spre{} \cat [\textbf{return}(t)]  \\
     & \whr \Spre{}, t = \TrE{E} \\
    \TrS{\textbf{return};} &= [\textbf{return}(\textbf{VoidType})] \\

    \TrS{\textit{ty}~\textit{id} = E;} &= \Spre{} \cat [\textbf{assign}(\textbf{local}, \textit{id}, t)] \\
     & \whr \Spre{}, t = \TrE{E} \\
    \TrS{\textit{ty}~\textit{id};} &= \emptylist \\

    \TrS{\textbf{throw}~E} &= \Spre{} \cat [\textbf{throw}] \\
    &   \whr \Spre{}, \_ = \TrE{E} \\

    \TrS{\textit{id} = E} &= \left\{\begin{array}{l l}
        \Spre{} \cat [\textbf{ref\_assign}(\textbf{read}(\mathcal{S}(\textit{id}), \textit{id}), t)] & \textbf{if} ~ \mathcal{R}(id)\\
        \Spre{} \cat [\textbf{assign}(\mathcal{S}(\textit{id}), \textit{id}, t)] & \textbf{otherwise} \\
    \end{array} \right.\\
    &   ~\whr \Spre{}, t = \TrE{E} \\

    \TrS{E.\textit{id} = E'} &= \Spre{} \cat \Spre{}' \cat [\textbf{ref\_assign}(t, t')] \\
     &\begin{array}{l l l}
        \whr & \Spre{}, t &= \TrE{E.\textit{id}} \\
          & \Spre{}', t' &= \TrE{E'} 
     \end{array} \\

    \TrS{E} &= \Spre{} \\
     & \whr \Spre{}, \_ = \TrE{E} \\

    \TrS{\textbf{if}(C)~\textit{St}_{\textit{then}}} 
     &= \TrS{\textbf{if}(C)~\textit{St}_{\textit{then}}~\textbf{else}~\{\}} \\
    \TrS{\textbf{if}(C)~\textit{St}_T~\textbf{else}~\textit{St}_F}
     &= \Spre{} \cat [\textbf{ite}(t, \TrS{\textit{St}_T}, \TrS{\textit{St}_F})] \\
     &\begin{array}{l l l}
        \whr & \Spre{}, t &= \TrE{C}\\
     \end{array} \\

    \TrS{\textbf{while}(C)~\textit{St}}
     &= \Spre{} \cat [\textbf{while}(t, S_B)] \cat [\textbf{flag}(\textbf{Break}, l_w)] \\
     &\begin{array}{l l l}
        \whr & l_w &\text{is a fresh label}  \\
        & S_B &= \TrS{\textit{St}} \cat [\textbf{flag}(\textbf{Continue}, l_w)] \cat \Spre{} \\
        & \Spre{}, t &= \TrE{C} \\
     \end{array} \\

    \TrS{\textbf{for}(\textit{St}_i ; C; \textit{It})~\textit{St}_B}
     &= \TrS{\textit{St}_i} \cat \Spre{c} \cat [\textbf{while}(t_c, S_B)] \\
     & ~~~ \cat [\textbf{flag}(\textbf{Break}, l_f)] \\
     &\begin{array}{l l l}
        \whr & l_f &\text{is a fresh label} \\
        & \Spre{c}, t_c &= \TrE{C} \\
        & S_B &= \mathbb{T}_S(\textit{St}_B, l_f) \cat \Spre{i} \\
        & & ~~~ \cat [\textbf{flag}(\textbf{Continue}, l_f)] \\
        & & ~~~\cat \Spre{c} \\
        & \Spre{i}, t_i &= \TrE{i}
     \end{array} \\

    \TrS{\textbf{continue};} &= [\textbf{jump}(\textbf{Continue}, l)]\\
    \TrS{\textbf{break};} &= [\textbf{jump}(\textbf{break}, l)] \\

    \TrS{\textbf{try}~\textit{St}~\textbf{catch}(...) \textit{St}_c} 
     &= \TrS{\textit{St}} \cat [\textbf{catch}(S_c)] \\
     & \whr S_c = \TrS{\textit{St}_c}
\end{array}$$

$$\begin{array}{l l}
\mathbb{T}_S \left(\begin{array}{l}
    \textbf{switch}(E_c) \{ \\
    \textbf{case} E_1 : S_1 \\
    \quad \vdots \\
    \textbf{case} E_n : S_n \\
    \textbf{default} : S_d \}
\end{array}, l \right)
&   = \begin{array}{l}
    \Spre{} \cat [\textbf{Assign}(\textbf{local}, \textit{id}, t_c)] \cat \\
    ~ [\textbf{ite}(c_1, \mathbb{T}_S(S_1, l') \cat \hdots \cat \mathbb{T}_S(S_n, l') \cat \mathbb{T}_S(S_d, l'), \emptylist)] \cat\\
    \quad \vdots \\
    ~ [\textbf{ite}(c_n, \mathbb{T}_S(S_n, l') \cat \mathbb{T}_S(S_d, l'), \emptylist)] \cat\\
    ~ \mathbb{T}_S(S_d, l') \cat [\textbf{Flag}(\textbf{Break}, l')] \\
    \end{array} \\
&   \begin{array}{l l l}
    \whr & l' & \textit{is a fresh label} \\
    &   \textit{id} & \textit{is a fresh variable ID}\\
    &   \Spre{c}, t_c &= \TrE{E_c} \\
    &   \_, t_i &=\TrE{E_i} ~\textbf{for}~i \in [1,n]\\
    &   c_i &= \textbf{equals}(\textbf{read}(\textbf{local}, \textit{id}), t_i)
    \end{array}
\end{array}$$

\end{definition}

We define the expression transformation rules in Definition \ref{def:TransE}, using the mapping $\mathbb{T}_E$.
The mapping produces a tuple, often denoted as $\Spre{}, t$.
The first tuple-argument consists of a, possibly empty, list of statements that should be executed beforehand, while the second tuple-argument is a term of the type $E$, see Figure \ref{fig:scpp_language}, which, upon evaluation, returns the value associated with the original expression.
This approach allows us to put any potential side-effects caused by the evaluation in the set of statements, e.g.\ as is done with the post-increment operator ($\textit{id}+\!\!+$).

We do not list the transformation rules for every binary and unary operator, as most of these follow very similar patterns, with the main difference being the resulting expression term, e.g.\ \textbf{Plus}.
Of note is the transformation rule for the logical and, i.e.\ $\&\&$, which can cause side-effects due to short-circuiting, i.e.\ if the first argument evaluates to false, the second argument is not evaluated, and thus no potential side-effects caused by said argument will occur.
The logical or, i.e.\ $||$, follows a similar pattern.

\begin{definition} \label{def:TransE}
The mapping $\mathbb{T}_E : \textit{Expression} \to \textit{List}(\textit{Scpp}) \times E$ is defined as follows, and maps \cpp expressions to a list of Scpp statements with side effects that are required to be executed beforehand, and an expression that yields the expected value, upon evaluation.

$$\begin{array}{l l l}
    \TrE{E_1 \&\& E_2} &= \Spre{1} \cat [\textbf{ite}(t_1, S_2, \textbf{assign}(\textbf{local}, \textit{id}'))], \textbf{read}(\textbf{local}, \textit{id}') \\
    &\begin{array}{l l l}
        \whr & \textit{id}' & \text{is a fresh variable ID}\\
          & S_2 &= \Spre{2} \cat \textbf{assign}(\textbf{local},t_2) \\
          & \Spre{1}, t_1 &= \TrE{E_1} \\
          & \Spre{2}, t_2 &= \TrE{E_2} 
    \end{array} \\

    \TrE{E_1 + E_2} &= \Spre{1} \cat \Spre{2}, \textbf{plus}(t_1, t_2) \\
     &\begin{array}{l l l}
        \whr &  \Spre{1}, t_1 &= \TrE{E_1} \\
          & \Spre{2}, t_2 &= \TrE{E_2} 
     \end{array} \\

    \TrE{! E} &= \Spre{}, \textbf{not}(t)\\
     & \whr \Spre{}, t = \TrE{E}\\

     \TrE{\textit{id}+\!\!+} &= [\textbf{assign}(\textbf{local}, \textit{id}', \textbf{read}(\mathcal{S}(\textit{id}), \textit{id}))] \cat S_{\textit{incr}}, \textbf{read}(\textbf{local}, \textit{id}')\\
      &\begin{array}{l l l}
        \whr & \textit{id}' & \text{is a fresh variable ID}\\
            & S_{\textit{incr}} &= \TrS{\textit{id} = \textit{id} + 1} \\
      \end{array} \\

    \TrE{E_c ? E_T : E_F} &= \Spre{c} \cat [\textbf{ite}(t_c, S_T, S_F)], \textbf{read}(\textbf{local}, \textit{id}')\\
    & \begin{array}{l l l}
        \whr & \textit{id}' & \text{is a fresh variable ID} \\
         & S_T &= \Spre{T} \cat [\textbf{assign}(\textbf{local}, \textit{id}', t_T)] \\
         & S_F &= \Spre{F} \cat [\textbf{assign}(\textbf{local}, \textit{id}', t_F)] \\
         & \Spre{c}, t_c &= \TrE{E_c} \\
         & \Spre{T}, t_T &= \TrE{E_T} \\
         & \Spre{F}, t_F &= \TrE{E_F} \\
    \end{array} \\
    
    \TrE{E . \textit{id}} &= \TrE{E \rightarrow \textit{id}} \\
    \TrE{E \to \textit{id}} &= \Spre{} \cat [\textbf{load}(\textit{id}', t, \textit{id})], \textbf{read}(\textbf{local}, \textit{id}') \\
     & \begin{array}{l l l}
        \whr & \textit{id}' & \text{is a fresh variable ID} \\
         & \Spre{}, t &= \TrE{E} 
     \end{array} \\
    
    \TrE{\textbf{this}} &= \emptylist, \textbf{self} \\

    \TrE{\textit{id}} 
    &   = \left\{ \begin{array}{l l}
        [\textbf{ref\_load}(\textit{id}', \textbf{read}(\mathcal{S}(\textit{id}), \textit{id})], \textbf{read}(\textbf{local}, \textit{id}') 
        &   \textbf{if} \mathcal{R}(\textit{id}) \\
        ~\whr \textit{id}' ~\text{is a fresh variable ID} \\
        \emptylist, \textbf{constant}(\textbf{EnumType}(\textit{id})) 
        &   \textbf{if}~\varepsilon(\textit{id}) \\
        \emptylist, \textbf{read}(\mathcal{S}(\textit{id}), \textit{id}) 
        &   \textbf{otherwise}\\
    \end{array} \right. \\

    \TrE{\textit{id} (E_1, \hdots, E_n)} 
    &   = \left\{\begin{array}{l l}
        \TrE{\textbf{this} \to \textit{id} (E_1, \hdots, E_n)} 
        &   \textbf{if}~\mathcal{F}(\textit{id})\\
        \Spre{} \cat [\textbf{callLambda}(\textit{id}', \textit{id}, \textit{ts}, \textit{refs})], \textbf{read}(\textit{local}, \textit{id}') 
        &   \textbf{otherwise}\\
        \whr \Spre{}, \textit{ts}, \textit{refs} = \mathbb{T}_A(\textit{id}, E_1, \hdots, E_n)
    \end{array}\right. \\

    \TrE{E \to \textit{id}_f (E_1, \hdots, E_n)} &= \Spre{o} \cat \Spre{} \cat [\textbf{call}(\textit{id}', t_o, \textit{id}_f, \textit{ts}, \textit{refs})], \textbf{read}(\textit{local}, \textit{id}') \\
     & \begin{array}{l l l}
        \whr & \textit{id}' & \text{is a fresh variable ID} \\
          & \Spre{o}, t_o &= \TrE{E} \\
          & \Spre{}, \textit{ts}, \textit{refs} &= \mathbb{T}_A(\textit{id}_f, E_1, \hdots, E_n)
     \end{array} \\

     \TrE{[\textit{cs}](\textit{ids}) \{ \textit{St} \} } 
      &= \emptylist, \textbf{init\_lambda}(\textit{ids}, \textit{copies}, \textit{refs}, S) \\
      & \begin{array}{l l l}
        \whr & l &\text{is a fresh label}  \\
           & S &= \TrS{\textit{St}} \\
           & \textit{copies} &= [c \in \textit{cs} | \mathcal{S}(c) = \textbf{local}] \\
           & \textit{refs} &= [c \in \textit{cs} | \mathcal{R}(c)]
      \end{array}
\end{array}$$
\end{definition}

\newpage

\begin{definition} \label{def:TransA}
The transformation mapping on function arguments $\mathbb{T}_A : \textit{ID}_f \times E^\mathbb{N} \to \textit{List}(\textit{Scpp}) \times \textit{List}(E) \times \textit{List}(\textit{ID})$ is defined as follows:

$$\begin{array}{l}
% main mapping
\begin{array}{l l}
\mathbb{T}_A(\textit{id}_f, \textit{arg}_1, \hdots, \textit{arg}_n) 
&   = \Spre{1} \cat \hdots \Spre{n}, [t_1, \hdots, t_n], \textit{id}_1 \cat \textit{id}_n \\
&   \whr \Spre{i}, t_i, \textit{id}_i = \mathbb{T}_A'(\textit{id}_f, i, \textit{arg}_i) \textbf{ for } i \in [1,n] 
\end{array} \\

% reference distinction
\begin{array}{l}
\mathbb{T}_A'(\textit{id}_f, i, E) = \left\{\begin{array}{l l}
    \mathbb{T}_A''(E) & \textbf{if}~\mathcal{R}_A(\textit{id}_f, i) \\
    \mathbb{T}_E(E) , \emptylist & \textbf{otherwise}
\end{array}\right. \end{array}\\

% pass-by-reference-mapping
\begin{array}{l l}
\mathbb{T}_A''(\textit{id})
&   = \left\{\begin{array}{ll}
    \emptylist, \textbf{c}\textbf{onstant}(\textbf{EnumType}(\textit{id})), \emptylist
    &   \textbf{if}~\varepsilon(\textit{id}) \\
    \emptylist, \textbf{read}(\textbf{local}, \textit{id}), [\textit{id}]
    &   \textbf{if}~\mathcal{R}(\textit{id}) \\
    \emptylist, \textbf{constant}(\textbf{LocalRef}(\textit{id})), [\textit{id}]
    &   \textbf{if}~\mathcal{S}(\textit{id}) = \textbf{local} \\
    \emptylist, \textbf{ref\_field}(\textbf{self}, \textit{id}), \emptylist
    &   \textbf{otherwise}
\end{array}\right.  \\

\mathbb{T}_A''(E \to \textit{id}) 
&   = \Spre{}' \cat [\textbf{assign}(\textbf{local}, \textit{id}, t)], \textbf{ref\_field}(\textbf{read}(\textbf{local}, \textit{id}')), \emptylist \\
&   \begin{array}{l l}
    \whr & \textit{id}' \text{ is a fresh variable ID} \\
    &   \Spre{}', t = \mathbb{T}_E(E)
    \end{array} \\

\mathbb{T}_A''(E)
&   = \mathbb{T}_E(E), \emptylist
\end{array} 
\end{array}$$
\end{definition}

\begin{figure}[h]
    \centering
    \includegraphics[width=0.5\linewidth]{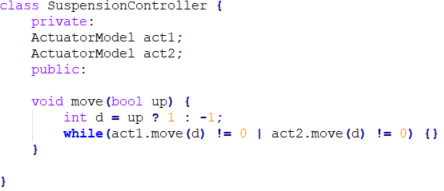}
    \caption{Example code that implements a moving platform.}
    \label{fig:suspension-controller}
\end{figure}

Let us discuss the transformation of part of a small example program, shown in Figure \ref{fig:suspension-controller}.
The suspension controller class controls a platform that can move up and down using two actuators, see Figure \ref{fig:actuator_model}.
The \textit{movePlatform} method raises the platform to either maximum or minimum heigth by fully extending or shrinking the actuators.
However it has to ensure that the length of the two actuators never differs by more than 1 or the platform would become too slanted.
This is done by repeatedly moving each actuator by only 1 till they no longer extend further.

When transforming the suspsension controller class, we populate the mappings in Definition \ref{def:semantics_stable}.
As such, the mapping $\textit{get\_prog}(\textit{raise})$ becomes as follows:
$$\begin{array}{l}
\textit{get\_prog(raise)} = [\\
\quad \begin{array}{l}      
    \textbf{ite}(\textbf{read}(\textbf{local}, \textit{up}),\\
    \quad [\textbf{assign}(\textbf{local}, \textit{freshVar0}, \textbf{constant}(\textbf{Number}(1)))],\\
    \quad [\textbf{assign}(\textbf{local}, \textit{freshVar0}, \textbf{constant}(\textbf{Number}(-1)))]), \\
    \textbf{assign}(\textbf{local}, d, \textbf{read}(\textbf{local}, \textit{freshVar0})), \\
    \textbf{call}(\textit{freshVar1}, \textbf{read}(\textbf{global}, \textit{act1}), \textit{grow}, [\textbf{read}(\textbf{local}, d)], \emptylist) , \\
    \textbf{call}(\textit{freshVar2}, \textbf{read}(\textbf{global}, \textit{act2}), \textit{grow}, [\textbf{read}(\textbf{local}, d)], \emptylist) , \\
    \!\!\!\begin{array}{l l}
    \textbf{while}(\textbf{or}( &\textbf{not\_equals}(\textbf{read}(\textbf{local}, \textit{freshVar1}), \textbf{constant}(\textbf{Number}(0))),\\
    &   \textbf{not\_equals}(\textbf{read}(\textbf{local},\textit{freshVar2}), \textbf{constant}(\textbf{Number}(0)))
    \end{array}), [ \\
    \quad \begin{array}{l}
        \textbf{flag}(\textbf{continue}, \textit{lbl3}) ,\\
        \textbf{call}(\textit{freshVar1}, \textbf{read}(\textbf{global}, \textit{act1}), \textit{grow}, [\textbf{read}(\textbf{local}, d)], \emptylist) , \\
        \textbf{call}(\textit{freshVar2}, \textbf{read}(\textbf{global}, \textit{act2}), \textit{grow}, [\textbf{read}(\textbf{local}, d)], \emptylist) ]) \\
        \end{array}, \\
    \textbf{flag}(\textbf{break}, \textit{lbl3})
    \end{array}\\
] 
\end{array}$$
\noindent The ternary operator (\_?\_:\_) is transformed into an \textbf{ite} statement and put in front of the assignment of $d$.
Similarly, the condition of the \textbf{while}-statement is replaced with reading the fresh variables \textit{freshVar1} and \textit{freshVar2}.
Subsequently, the method calls, now assigning their results to the new fresh variables, are put before and inside of the \textit{while} loop.

Besides the \textit{get\_prog} mapping, the $i_{\stloc}$ mapping, which is responsible for initializing the parameter variables, is populated as follows:
$$i_{\stloc}(\textit{movePlatform}, \textit{vs}) = \lambda \textit{id}_x. \textbf{VoidType}[\textit{up} \mapsto vs . 0]$$

Besides this the data types for identifiers and flag labels are also populated.
The variable identifiers is populated with the member names for the actuators, the local variables inside the \textit{movePlatform} method and the fresh variables are added as variable identifiers.
Both the SuspensionControllers own \textit{movePlatform} method and the called Actuators \textit{move} method are added to $\textit{ID}_f$.
The identifiers, used to distinguish the various processes, are updated with an identifier for the main SuspensionControllers, and identifiers for its two members.
The set of flag labels is populated with the fresh label used for the while loop.
\textbf{while}
$$\begin{array}{l}
\textit{ID}_x := \textit{up} | \textit{d} | \textit{act1} | \textit{act2} | \textit{freshVar0} | \textit{freshVar1} | \textit{freshVar2}.\\
\textit{ID}_f := \textit{movePlatform} | \textit{move} .\\
\textit{ID}_p := \textit{SuspensionController} | \textit{ActuatorModel1} | \textit{ActuatorModel2} .\\

\textit{Lbl} := \textit{lbl3} .\\
\end{array}$$

At last the final composition of processes, used to generate the model is as follows:
$$\begin{array}{l}
\textbf{allow}(\{\texttt{call\_func}, \texttt{return\_func}, \texttt{throw\_func}, \texttt{load\_comm}, \texttt{store\_comm}\}, \\
\textbf{comm}(\{ \texttt{call\_func\_t} | \texttt{call\_func\_b} \to \texttt{call\_func}, \\
\tab{}{1}\texttt{return\_func\_t} | \texttt{return\_func\_b} \to \texttt{return\_func}, \\
\tab{}{1}\texttt{throw\_func\_t} | \texttt{throw\_func\_b} \to \texttt{throw\_func},\\
\tab{}{1} \texttt{load\_t} | \texttt{load\_b} \to \texttt{load\_comm}, \\
\tab{}{1} \texttt{store\_t} | \texttt{store\_b} \to \texttt{store\_comm}\}, \\
\tab{}{2}     \textit{TopInterface} ~||~ P'(\textit{ActuatorModel1}) ~||~ P''(\textit{ActuatorModel2}) ~||\\
\tab{}{2}\begin{array}{@{} l @{} l}
         C(\textit{SuspensionController}, \lambda \textit{id}_x. \textbf{VoidType} [& \textit{act1} \mapsto \textbf{PType}(\textit{ActuatorModel1}), \\
         & \textit{act2} \mapsto \textbf{PType}(\textit{ActuatorModel2})]) ~))\\
    \end{array} \\
\end{array}$$
Here, \textit{TopInterface}, described in the next section, is a process that triggers the actual calls to the SuspensionController.
The processes $P'$ and $P''$ are the processes for the two actuator models, which can be constructed using the transformation tool or any of the other methods that are described in the next section.
Finally, the process $C$ is the stable process and is passed the identifier for the SuspensionController and the initial mapping of its global variables.
As such its members \textit{act1}, and \textit{act2} are initialized with the respective process identifiers of the member processes.

\section{The SSTraGen tool} \label{sec:sstragen}
The GUI-based State Space Transformation and Generation tool, SSTraGen for short, implements our discussed framework.
The tool generates a process equation in mCRL2.
Both the source code as well as a precompiled version of the tool are publicly available online \cite{repo}.
Both versions include a small set of example programs, including the examples shown thus far.
Running the examples to their full extent, requires the mCRL2 toolset \cite{mcrl2toolset}.
As we discuss in Section \ref{sec:conclusion}, the public version does not offer support for multi-file projects and requires a single c++ file with all declarations or at least forward declarations.

The tool offers a graphical user interface for transforming c++ files, setting bounds, and providing abstractions for sub-components.
Different configurations can be easily swapped around and can be saved and loaded from a project file.
Besides the GUI based tool, we also provide a simple command line interface that allows for rerunning the transformation process whilst keeping the current configuration of user input.
This can be used in cases in which a code change has occured and we are interested in acquiring a model of the new code.
We outline a common use case for this later in this section.

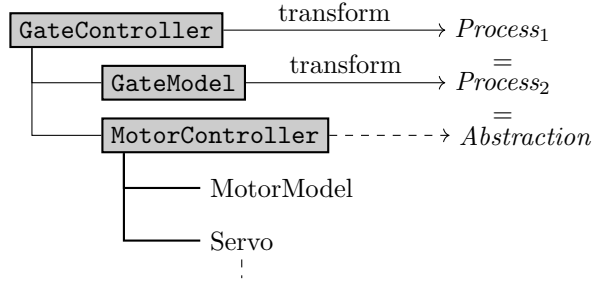
\begin{figure}[h]
    \centering
    \tikzstyle{treenode}=[draw=black,thick,fill=gray!40, anchor=west]
    \begin{tikzpicture}[%
      grow via three points={one child at (-.2,-0.7) and
      two children at (-.2,-0.7) and (-.2,-1.4)},
      edge from parent path={(\tikzparentnode.south west) ++(0.3cm, 0) |- (\tikzchildnode.west)}]
    \node[treenode] (GateController) {\texttt{GateController}}
    child { node[treenode] (GateModel) {\texttt{GateModel}}}
    child { node[treenode] (MotorController) {\texttt{MotorController}}
        child[treenode] { node {MotorModel}}
        child[treenode] { node (Servo) {Servo}}
    };

    \draw[dashed] (Servo) -- ++(0,-.5cm);
    %\draw[decorate,decoration={brace,amplitude=10pt}] (a1.north west) -- (a99.south west) node [midway,xshift=-0.4cm,left] {Section foo};
`   
    \node[right = 2.95cm of GateController](p1) {$\textit{Process}_1$};
    \node[right = 2.65cm of GateModel] (p2) {$\textit{Process}_2$};
    \node[right = 1.6cm of MotorController] (p3) {\textit{Abstraction}};

    \draw[->] (GateController) -- node[midway, above] {transform} (p1);
    \draw[->] (GateModel) -- node[midway, above] {transform} (p2);
    \draw[dashed, ->] (MotorController) -- (p3);

    \node[below= 0 of p1] {=};
    \node[below= 0 of p2] {=};
    
    \end{tikzpicture}
    \caption{By using an abstraction, we avoid transforming multiple sub-components.\\
    The final process consists of the parallel composition of the two processes acquired through transformation and the abstraction process.}
    \label{fig:component-wise-transformation}
\end{figure}

SSTraGen allows for component-wise transformation of classes.
In other words, the programmer decides what component and subcomponents should be transformed directly and what subcomponents should be abstracted over.
Take the transformation shown in Figure \ref{fig:component-wise-transformation} as an example.
In this example, the main component, i.e.\  \texttt{GateController}, has two members as subcomponents: \texttt{GateModel} and \texttt{MotorController}; the latter, in turn has two subcomponents on its own, which may in turn have their own subcomponents.
As transforming MotorController and all of its (indirect) subcomponents is undesirable, the software engineer instead decides to provide an abstraction of the MotorController.
The resulting end process will be the parallel composition of the TopInterface, which we explain later, the transformations of GateController and GateModel, and the abstraction of the MotorController.

We allow for two types of model abstractions: custom processes, and bottom stubs.
Custom processes are handwritten processes, which can be useful for modelling systems with state based behavior, e.g.\ locks.
Bottom stubs are stateless processes that implement over-approximations of the given class interface, i.e.\ return values are (at most) bounded.

Bottom stubs are of the following form:
$$\begin{array}{l}
    \textit{Stub}(\pid : \textit{ID}_p) = \\
    \quad \sum_{\fid {:} \textit{ID}_f, \textit{vs}{:}\textit{List}(V), \textit{lras}{:}\textit{List}(\textit{ID}_x \times V)}. \texttt{call\_func\_b}(\pid, \fid, \textit{vs}, \textit{lras}) \cdot ( \\
    \begin{array}{l l}
        & \sum_{v{:}V}.(\textit{is\_within\_bounds}(\fid, v)) \to \texttt{return\_func\_b}(\pid, \fid, v, \textit{lras}) \cdot \textit{Stub}(\pid)\\
    +   & (\textit{can\_throw}(\fid)) \to \texttt{throw\_func\_b}(\pid, \fid, \textit{lras}) \cdot \textit{Stub}(\pid) ) 
    \end{array} \\
    +  \sum_{\textit{id}_x{:}\textit{ID}_x, v{:}V}. \texttt{store\_b}(\pid, \textit{id}_x, v) \cdot \textit{Stub}(\pid) \\
    +   \sum_{\textit{id}_x{:}\textit{ID}_x, v{:}V}. (\textit{is\_within\_bounds}(\textit{id}_x, v)) \to \texttt{load\_b}(\pid, \textit{id}_x, v) \cdot \textit{Stub}(\pid)
\end{array}$$
\noindent Here, \textit{is\_within\_bounds} evaluates to true iff the given value(s) are of valid type and fall within any potential user-set bounds, and \textit{can\_throw} evaluates to true iff the given method is expected to be able to (sometimes) throw and exception.

The external usage of our subject class is directed by the so-called top-level interface.
The top-level interface is a simple process that can call any of the various methods of the subject class, and/or access any of its fields.
The parameters for the method calls and values assigned to any of the fields of the subject are bounded for the same reason as the return values on bottom stubs.

Top-level interfaces are of the following form:
$$\begin{array}{l}
 \textit{TopInterface}(\pid : \textit{ID}_p) = \\
 \quad \sum_{\fid{:}\textit{ID}_f}. (\textit{is\_visible}(\fid)) \to \sum_{\textit{vs}{:}\textit{List}(V)}. (\textit{is\_within\_bounds}(\fid, \textit{vs})) \to \\
 \quad \quad \texttt{call\_func\_t}(\pid, \fid, a(\fid, \textit{vs}), l(\fid, \textit{vs})) \cdot (\\
 \quad \quad \begin{array}{l l}
      & \sum_{v{:}V, \textit{lras}:\textit{List}(\textit{ID} \times V)}. \texttt{return\_func\_t}(\pid, \fid, v, \textit{lras}) \cdot \textit{TopInterface}(\pid)\\
    + & \sum_{\textit{lras}:\textit{List}(\textit{ID} \times V)}\texttt{throw\_func\_t}(\pid, \fid, \textit{lras}) \cdot \\
      & ((\neg \textit{throws\_terminates}(\fid)) \to \textit{TopInterface}(\pid)))
 \end{array} \\
 +  \sum_{\textit{id}_x{:}\textit{ID}_x}. (\textit{is\_visible}(\textit{id}_x)) \to \sum_{v{:}V}. \texttt{load\_t}(\pid, \textit{id}_x, v) \cdot \textit{TopInterface}(\pid) \\
 +  \sum_{\textit{id}_x{:}\textit{ID}_x}. (\textit{is\_visible}(\textit{id}_x)) \to \sum_{v{:}V}. (\textit{is\_within\_bounds}(\textit{id}_x, v)) \to \\
 \quad \quad \texttt{store\_t}(\pid, \textit{id}_x, v) \cdot \textit{TopInterface}(\pid)
\end{array}$$

\noindent Here, \textit{is\_visible} indicates whether or not a given field is externally visible/accessible, and \textit{is\_within\_bounds} is the same as for the bottom stub, and the \textit{throws\_terminates} mapping indicates whether the given system should halt if the given function throws an exception.
These are generally inferred from the transformed code, but can also be configured by the user.

\subsection{Bounds}
\begin{figure}
    \centering
    \includegraphics[width=0.9\linewidth]{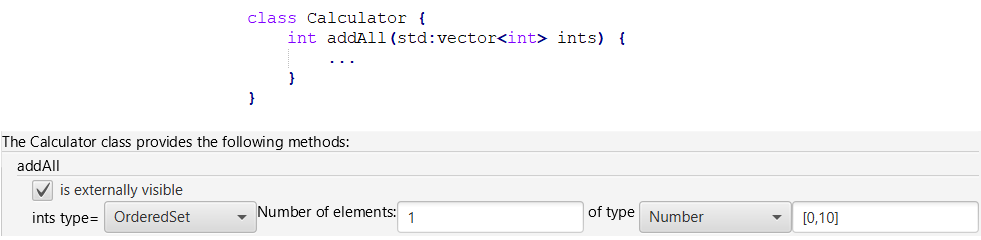}
    \caption{Through the GUI, the user can bound the amount of possible arguments.
    In this case only lists with exactly one integer, ranging from 0 up till and including 10, as an element.}
    \label{fig:screenshot_bounds}
\end{figure}

The bounding of top level arguments and return values at the bottom, i.e.\ configuring the implementation of \textit{is\_within\_bounds}, is done via the tool, see Figure \ref{fig:screenshot_bounds}.
Every parameter comes with a so called BoundedType, a tuple consisting of a ValueType and a possible bound.
ValueTypes are a subset of the possible constructors of the data sorts in $V$, see Figure \ref{fig:scpp_language}, specifically: Number, Boolean, VoidType, OrderedSet, PType, EnumType, and StringType.
The type of bound differs per ValueType.
Besides these, a BoundedType can also be an UnknownType for cases in which the transformer could not infer the type of a given argument or return value.
In such cases the user needs to provide the type as input themselves before a model can be generated.

Number bounds allow us to bound any type of number, e.g. \texttt{int}, \texttt{float}, \texttt{long}.
The bounds can be a combination of ranges and/or constants.
Ranges can also be specified with an interval, this allows us to keep ranges over types such as floats finite.

Boolean bounds are straightforward, and can be set to either always be \texttt{true} and/or \texttt{false}.
EnumTypes are unbounded besides ensuring that only elements of that specific enum are allowed.
For OrdereredSet bounds, we can set the size of the set, and a BoundedType on the elements of the set, as is shown in the example figure.

PType bounds, i.e.\ bounds over an argument/return value that is a class, make use of so called dynamic categories.
Dynamic categories are collections of various implementations, i.e.\ processes acquired through either transformation or abstraction, of a given class/interface, often in the form of various stubs.
PType bounds can be set to a specific category; the argument/return value can then be a processes id of any process in the given category. 
Any element of the involved dynamic categories is parallel composed with the other processes such that it can be interacted with by using the associated process id, in the same way one would call the methods or access the fields of one of its members, i.e.\ through the actions in Definition \ref{def:action_labels}.

\subsection{Global namespace}
A common, albeit sometimes shunned, practice within OOP is the usage of globals, i.e.\ fields that can be accessed from anywhere in the code.
To be able to support this, we add one additional process, the \textit{GlobalState}, to our composition of processes.
The \textit{GlobalState} process has a unique process identifier $\textit{GlobalNamespace} \in \textit{ID}_p$, so that we can access global fields and methods using our standard actions, e.g.\ $\sum_{v{:}V}.\texttt{load\_t}(\textit{GlobalNamespace}, \textit{field}, v)$ where \textit{field} is the name of the global field.
This global process, can either be a custom process, or a bottom stub, such that we can control what the expected outcome is of accessing static methods and fields.

\subsection{Command Line Interface}
Finding the right levels of abstractions and bounds forms the main bottleneck of this approach, as it is the one aspect that is not fully automated.
To alleviate this somewhat, the SSTraGen tool allows for the reusing of user input when applied to different software iterations of the same component such that the transformation of code into an equivalent model can be fully automated.
A given configuration can be stored in memory as a project file, using the tool.
This project file contains the user input, configuration of processes, any abstraction models, and the load instructions for each associated translation unit.

Through a command line interface or the GUI, the user can use a project file to redo the transformation on any translation unit that might have since been changed. 
After this transformation occurs, the new processes are used to generate a potentially different model, which can be very useful during software development.
For example, when a pull request is made, a model of the code can now be automatically generated.
This new model can then be compared with the previous model by testing for equivalence modulo some equivalence relationship, e.g.\ weak-trace equivalence, to see if any regression occurred.
Or, instead of using tests, much more expressive modal formulas can be verified on the generated model, to see if the code changes do not invalidate any requirements.

\section{Case study} \label{sec:case-study}
We have used the SSTraGen tool on several software components within Philips IGT; both to showcase the usefulness of acquiring models from actual code and to test whether the tool can actually handle the less-than-ideal nature of industrial code.
Applying the tool to the various components was very much an iterative process, in which many of the obstacles encountered actually led to improvements to the tool and the previously discussed semantics and transformation rules.
Eventually, the tool was able to successfully transform the object-oriented code present and create an accurate model for every component.
The accuracy of the models was verified in two ways: by discussing the model with the programmers maintaining the code to see if the model captured any behavior that the programmer did not expect, and by transforming software tests into modal formulas and verifying whether the formulas hold on our model,
We also repeated the last step for incorrect versions of the existing tests, e.g.\ test for incorrect return values.
It should be noted that modal formulas are more powerful than tests, as tests can only test whether a requirement holds in a given execution, whereas modal formulas can test if a requirement holds in any given state, or any given state satisfying some property.

We now discuss the application of our tool on two of the components within Philips IGT.
The first component is relatively small, consisting of just over 600 lines of code.
However, what makes this component interesting is the complexity introduced by its members/sub-components.
The given component effectively consists of two fixed-size lists of sub-components.
Both types of sub-components, function as a type of lock that can be in either a locked or unlocked state, and have some sort of ``lock" and ``unlock" function to change state, in which the ``lock" function succeeds, i.e. returns true, iff the specific sub-component is ``unlocked".
To make matters even more complex, one type of sub-component can also fail indefinitely, i.e. once an exception is thrown by one of its functions, any function call on that sub-component after that will also throw an exception.

We extracted a model of the component for lists with $1,2,$ and $3$ of each sub-component.
We first used action hiding on each model to ensure that only the calls to the main component are visible.
This exposes the interface of the component, i.e.\ what return value can be expected at any given moment/state.
We expect the component to have approximately $2^n \times 3^n$ discernable stable states, where $n$ is the size of the two respective lists of subcomponents.
However, applying action hiding on the models leads us to an interesting observation which is that, regardless of the list size, the interface only has $3$ discernable states.

Inspecting the internal states of the model reveals that the component ensures that each sub-component is always either "locked", or "unlocked", till at least one component has failed, in which case, it will always propagate the failure, regardless of which sub-component failed.
This observation was verified again by rephrasing it as a modal formula, and proving that the formula holds on each model.
This observation is noteworthy as it showcases how the tool and other model-based techniques can be used to gain additional understanding of complex software systems in a way that is often difficult through other techniques.
For example, models acquired through learning would not allow for the inspection of the internal states as this is not observable behavior.

The second component is a much larger component, consisting of over 1000 lines of code.
To be precise, the implementation of the component/class that we are interested in is split up over multiple files. The implementation file containing the behavior that we are interested in / have transformed plus header is just over 1000 lines of code alone; everything combined is much larger.
This component, which we will simply refer to as the \textit{Adapter}, oversees the communication between two important parts of the live x-ray imaging machine. 
As this communication is critical, the software engineers responsible have formalized a functional view of the communication occurring between the two halves of the machine in the form of a UML state diagram of approximately 15 states.
The Adapter itself is responsible for receiving signals from either side and then triggering methods belonging to the other side.
Which methods, if any at all, need to be triggered is dependent on the incoming signal and the current state.
Adding to the complexity of the system is that communication between the two halves happens under distinct semantics depending on the direction of the signal being sent.

With the mentioned complexity in mind, it is apparent that it is difficult for a programmer or software engineer to see from the actual code as to whether or not it conforms to the UML diagram.
As such, we set out to see if the SSTraGen tool could be used to acquire a model of the Adapter code and use model based techniques to answer the question as to whether or not it conforms to the UML diagram.
After transforming the code and providing the proper abstraction for all member classes involved, and a small amount of manually fixing the generated mCRL2 code, we were able to generate the statespace of the software component.

To make the model easier to work with, both from a user perspective as well as to ensure to reduce the run-time of any algorithms required, we reduce the statespace of the model modulo divergence preserving branching bisimulation.
This reduction preserves many important aspects of the original model, such as deadlock and livelock \cite{luttik}.
Before statespace reduction, the model contains $19.6k$ states and $20.6k$ transitions.
After statespace reduction, the model contains only $936$ states and $1368$ transitions.
The model showed to be a clear subset of the UML state diagrams allowed behavior.

The process of transformation and providing correct abstractions for the members of the component took about half a workdays of time, with almost all of the time being in the latter.
When discussing the model with the software engineer responsible, it was noted that something like this would usually take about a workweek worth of time to manually read through the code to gain the same perspective.
Although, this is clearly purely anecdotal, we believe that this showcases that there are evidently use cases within the software engineering industry in which transforming code into a behavioral model in such a way can prove to be very beneficial.

\section{Conclusion \& Future Work} \label{sec:conclusion}
We have outlined a framework for transforming object oriented code, particularly \cpp, into equivalent behavioral models.
We have created a GUI based tool that implements said framework, and have shown that this tool can be successfully applied to the large scale of complexity encountered within real-world software systems.
Whilst we do not think that this is the fits-all solution for acquiring models from real worlds systems, we do believe that, through our case studies at Philips, we have shown that there is an important subset of software systems in which this is useful.

We now shortly discuss the practical limitations of our tool, the reason behind these, and potential solutions to these shortcomings.

First, we note that, while the framework can be easily applied to other OOP languages, the current version of our tool is specific to \cpp.
Due to the larger context of the research project within which this tool and research was done, it was decided to apply our framework to \cpp, as this is mostly the language used within Philips IGT.
As correctly parsing \cpp is quite difficult, we decided to use an already existing parser which provides access to the AST; for this we decided to use the Eclipse CDT parser.
The Eclipse CDT parser, being the tools main dependency, brings its own limitation, namely: Eclipse CDT only supports up to \cpp version 17, and the parser can not parse partial translation units.

Because we require access to the entire translation unit, i.e.\ including the imports, we make use of a proprietary extension of Eclipse CDT, developed by TNO ESI \cite{renaissance}, specifically aimed at supporting the build architecture used within Philips IGT.
As such, the public version of the tool requires the user to manually add forward declarations to their translation units to ensure that they can be used by the tool.

A potential solution to both these problems would be to develop transformation rules for the LLVM intermediate representation language \cite{llvm-ir}.
This would mean that any language with a known LLVM transpiler would be supported.
However, we note that big companies will often have their own custom build scripts to deal with the complex nature of their systems, and as such might always require a custom solution for acquiring a full translation unit.

Besides these, we also highlight some possible future work:
\begin{itemize}
    \item With our novel transformation and specification framework for object oriented classes in mind, we can now look at what changes to code mean. Particularly, how do we specify a change to a class that will cause detectable behavioral changes, and how do we test equivalence of classes modulo said behavioral changes.
    \item A strong limitation in our approach is the limitation to a finite amount of processes. 
    Because of this limitation, models of dynamically allocated instances must be forced to have a single state.
    Proper solutions to this, would in turn aid in the expressivity of approaches such as this.
    \item Finally, towards increasing expressivity, the SCPP semantics for processing states could be extended such that it no longer is forcd to run under run-to-completion semantics.
    However, it should be noted that this generally will lead to a statespace explosion within the model due to the arbitrary interleaving of multiple function calls.
\end{itemize}

\bibliographystyle{llncs/splncs04}
\bibliography{bibliography}

\newpage
\begin{appendix}
\section{Process Semantics} \label{sec:appendixA}

\begin{definition} \label{def:semantics_Scpp}
    The define the process equation $P$ for processing states $\textit{State} = \textit{ID}_p \times V^{\textit{ID}_x} \times \textit{Stack}$ as follows:

{\fontsize{6}{8}\selectfont 
$
\begin{array}{l}
% begin P process equation
P(\st = \langle \pid : \ID, \stglob : \State, \textit{call\_stack} {:} \textit{List(StackFrame)} \rangle) = \textbf{match}~\textit{call\_stack}~\textbf{with} \\
    \tab{}{1} \STF{\textit{prog} : \textit{List(S)}}{x : \textit{ID}}{\stloc}{\textit{refs} : \textit{List}(\textit{ID} \times V)}{stl} \to (\textit{prog} \approx [])\\
    \tab{}{2} \to P_{\textit{return}}(\textbf{VoidType}, \st) \\
    \tab{}{2} \diamond ~\textbf{match} ~ \textit{hd(prog)} ~ \textbf{with}\\
    % begin individual scpp statement semantics
\tab{}{3}\begin{array}{l c l}
    \textbf{return}(e) & \to & P_{\textit{return}}(\semE(e,\st), \st) \\
    
    \textbf{assign}(\textbf{local}, y, e) & \to &  \tau \cdot P(\pid, \stglob, \STF{\tlprog}{x}{\stloc[y \mapsto \semE(e, \st)]}{\textit{refs}}{stl}) \\
    
    \textbf{assign}(\textbf{global}, y, e) & \to & \tau \cdot P(\pid, \stglob[y \mapsto \semE(e, \st)], \STF{\tlprog}{x}{\stloc}{\textit{refs}}{stl}) \\
    
    \textbf{ite}(c, \textit{prog}_T, \textit{prog}_F) & \to & (\semE(c, \st)) \\
    & & \quad \begin{array}{c l}
        \to & \tau \cdot P(\pid, \stglob, \STF{\textit{prog}_T \cat \tlprog}{x}{\stloc}{\textit{refs}}{stl})\\
        \diamond & \tau \cdot P(\pid, \stglob, \STF{\textit{prog}_F \cat \tlprog}{x}{\stloc}{\textit{refs}}{stl})\\
        \end{array}\\

    \textbf{while}(c, \textit{prog}_W) & \to & (\semE(c, \st)) \\
    & & \quad \begin{array}{c l}
        \to & \tau \cdot P(\pid, \stglob, \STF{\textit{prog}_W \cat \textit{prog}}{x}{\stloc}{\textit{refs}}{stl})\\
        \diamond & \tau \cdot P(\pid, \stglob, \textit{nextFrame})\\
        \end{array}\\

    \textbf{call}(y, e_p, \fid, \textit{es}, \textit{refs}) & \to & \textbf{match}~\semE(e_p, \st)~\textbf{with}\\
    \multicolumn{3}{l}{\tab{}{2}\begin{array}{l l l @{} l}
            \textbf{PType}(\pid) & \to &
            \tau \cdot P(\pid, \stglob, \textbf{frame}( & \textit{get\_prog}(\fid), y, \\
            & & & i_{\stloc} (\fid, \semE^*(\textit{es}, \st)), \\
            & & & \textit{makeLRAs}(\textit{refs}, \stloc)) {::} \textit{nextFrame}) \\
            \textbf{PType}(\pid') & \to & \multicolumn{2}{l}{\texttt{call\_func\_t}(\pid', \fid, \semE^*(\textit{es}, \st), \textit{makeLRAs}(\textit{refs}, \stloc)) \cdot ( }\\
            & & \multicolumn{2}{l}{\quad \sum_{v{:}V, \textit{lras}{:}\textit{List}(\textit{ID}_x \times V)}. \texttt{return\_func\_t}(\pid', \fid, v, \textit{lras}) \cdot}\\
            & & \quad P(\pid, \stglob, \textbf{frame}( & \tlprog, x,\\
            & & & \textit{ref\_refs}_\sigma(\textit{lras}, \stloc[y \mapsto v]),\\
            & & & \textit{ref\_refs}_r(\textit{lras}, \textit{refs})) {::} \textit{stl})\\
            & & \multicolumn{2}{l}{+ \sum_{\textit{lras}:\textit{List}(\textit{ID}_x \times V)}. \texttt{throw\_func\_t}(\pid', \fid, \textit{lras}) \cdot } \\
            & & \multicolumn{2}{l}{\begin{array}{l @{} l}
                P_{\textit{throw}}(\pid, \stglob, \textbf{frame}(& \tlprog, x, \\
                 & \textit{ret\_refs}_\sigma(\textit{lras}, \stloc), \\
                & \textit{refs}) {::} \textit{stl}))
            \end{array}} 
        \end{array}} \\
    
    \textbf{throw} & \to & 
        \tau \cdot P_{\textit{throw}}(\st) \\

    \textbf{catch}(\textit{prog'}) & \to &
        \tau \cdot
        P(\pid, \stglob, \textit{nextFrame}) \\

    \textbf{ref\_load}(\textit{to}, e) & \to & \textbf{match}~\semE(e, \st)~\textbf{with}\\
    \multicolumn{3}{l}{\tab{}{2}\begin{array}{l c l}
        \textbf{FieldRef}(\pid', \textit{id}_x') & \to & 
        \sum_{v{:}V}. load\_t(\pid', \textit{id}_x', v) \cdot \\
        & & \quad P(\pid, \stglob, \STF{\tlprog}{x}{\stloc[\textit{id}_x \mapsto v]}{\textit{refs}}{stl})\\
        \textbf{LocalRef}(\textit{id}_x) & \to &
        \tau \cdot P(\pid, \stglob, \STF{\tlprog}{x}{\stloc[\textit{to} \mapsto \textit{refE}(\textit{id}_x, \textit{refs})]}{\textit{refs}}{\textit{stl}})
    \end{array}}\\

    \textbf{ref\_assign}(e_{\textit{to}}, e) & \to & \textbf{match}~\semE(e_{\textit{to}}, \st)~\textbf{with}\\
    \multicolumn{3}{l}{\tab{}{2}\begin{array}{l c l}
        \textbf{FieldRef}(\pid ', \textit{id}_x') & \to & 
        store\_t(\pid', \textit{id}_x', \semE(E', e')) \cdot
        P(\pid, \stglob, \textit{nextFrame}) \\
        \textbf{LocalRef}(\textit{id}_x') & \to &
        \tau \cdot P(\pid, \stglob, \STF{\tlprog}{x}{\stloc}{\textit{updateLRA}(\textit{id}_x', \semE(e, \st), \textit{refs})}{\textit{stl}}\\ 
    \end{array}}\\

    \textbf{jump}(\textit{flag}) & \to & 
        \tau \cdot
        P(\pid, \stglob, \STF{\textit{jump\_to}(\textit{flag}, \tlprog)}{x}{\stloc}{\textit{refs}}{stl})\\

    \textbf{flag}(\textit{flag}) & \to &
        \tau \cdot
        P(\pid, \stglob, \textit{nextFrame}) \\

    \multicolumn{3}{l}{\textbf{call\_lambda}(y, \textit{id}_\lambda, [\textit{arg}_1, \hdots, \textit{arg}_n], \textit{refs}) \to \textbf{match}~\stloc (\textit{id}_\lambda)~\textbf{with}}\\
    \multicolumn{3}{l}{\tab{}{2}\textbf{LambdaType}([\textit{par}_1, \hdots, \textit{par}_n], \textit{lras}, [\textit{ref}_1, \hdots, \textit{ref}_m], \textit{prog}') \to}\\
    \multicolumn{3}{l}{\tab{}{3}\begin{array}{l l l}
        \tau \cdot P(\pid, \stglob, \textbf{frame}(\!\!\! & \textit{prog}', y & \\
         & \textit{ret\_refs}(\textit{lras}, \stloc \!\!\! & [\textit{par}_1 \mapsto \semE(\textit{arg}_1, \st)] \hdots [\textit{par}_n \mapsto \semE(\textit{arg}_n, \st)] \\
         & & [\textit{ref}_1 \mapsto \textbf{LocalRef}(\textit{ref}_1)] \hdots [\textit{ref}_m \mapsto \textbf{LocalRef}(\textit{ref}_m)]), \\
         & \multicolumn{2}{l}{\textit{make\_LRAs} ( [\textit{ref}_1, \hdots, \textit{ref}_m \cat \textit{refs}, \stloc) )) :: \textit{nextFrame}}
        \end{array}} \\
    
    \end{array} \\
\text{where} \\
P_{\textit{return}}(v:V, \st = \langle \pid : \ID, \stglob : \State, \textit{call\_stack} {:} \textit{List(StackFrame)} \rangle) = \textbf{match}~\textit{stl}~\textbf{with}\\
    \quad \begin{array}{l c l}
        \textbf{external}~f_{\textit{id}} & \to & \texttt{return\_func\_b}(f_{\textit{id}}, v, \textit{refs}) \cdot C(\pid, \stglob)  \\
        \STF{\textit{prog}'}{x'}{\stloc'}{\textit{refs}'}{stl'} & \to & \tau \cdot P(\pid, \stglob, \STF{\textit{prog}'}{x'}{\textit{ret\_refs}(\textit{refs}, \stloc'[x \mapsto v])}{\textit{refs}'}{stl'})

    \end{array} \\

P_{\textit{throw}}(\st = \langle \pid, \stglob, \textit{call\_stack}\rangle) = \textbf{match}~\textit{call\_stack}~\textbf{with}\\
    \quad \STF{\textit{prog}}{x}{\stloc}{\textit{refs}}{\textit{stl}} \to \textbf{match}~\textit{resolve\_throw}(\textit{prog})~\textbf{with}\\
    \quad \quad \begin{array}{l c l}
        \textbf{framebreak} & \to & 
        \textbf{match}~\textit{stl}~\textbf{with}\\
         & & \begin{array}{l c l}
            \textbf{external}~\fid & \to &
            \texttt{throw\_func\_b}(\pid, \fid, \textit{refs}) \cdot C(\pid, \stglob)\\
            \STF{\textit{prog}'}{x'}{\stloc'}{\textit{refs}'}{\textit{stl}'} & \to &
            \tau \cdot P_{\textit{throw}}(\pid, \stglob, \STF{\textit{prog}'}{x'}{\stloc'}{\textit{refs}'}{\textit{stl}'})
         \end{array} \\
        
        \textbf{caught}(\textit{prog}') & \to & 
        \tau \cdot P(\pid, \stglob, \STF{\textit{prog}' \cat \tlprog}{x}{\stloc}{\textit{refs}}{\textit{stl}})
        \end{array} \\
        
\textit{resolve\_throw}(\emptylist) = \textbf{framebreak} \\
\textit{resolve\_throw}(p {::} \textit{prog}) = \left\{\begin{array}{l l}
    \textbf{caught}(\textit{prog}') & \textbf{if}~p = \textbf{catch}(\textit{prog}')\\
    \textit{resolve\_throw}(\textit{prog}) & \textbf{otherwise}
\end{array}\right. \\

\textit{jump\_to}(\textit{flag}, \textit{prog}) = \left\{ \begin{array}{l l}
    \tlprog & \textbf{if}~\textit{hd}(\textit{prog}) = \textit{flag} \\
    \textit{jump\_to}(\textit{flag}, \tlprog) & \textbf{otherwise}
\end{array}\right.

\end{array}
$
}

\end{definition}

\section{Implementation Details of the SSTraGen Tool}
The following section of the Appendix contains diagrams and descriptions regarding the design and precise implementation of the SSTraGen tool.
The information presented in this section is implementation-specific and does not form a requirement for implementing the framework outlined in the main text.

\subsection{Language generic transformation process}
\begin{figure}
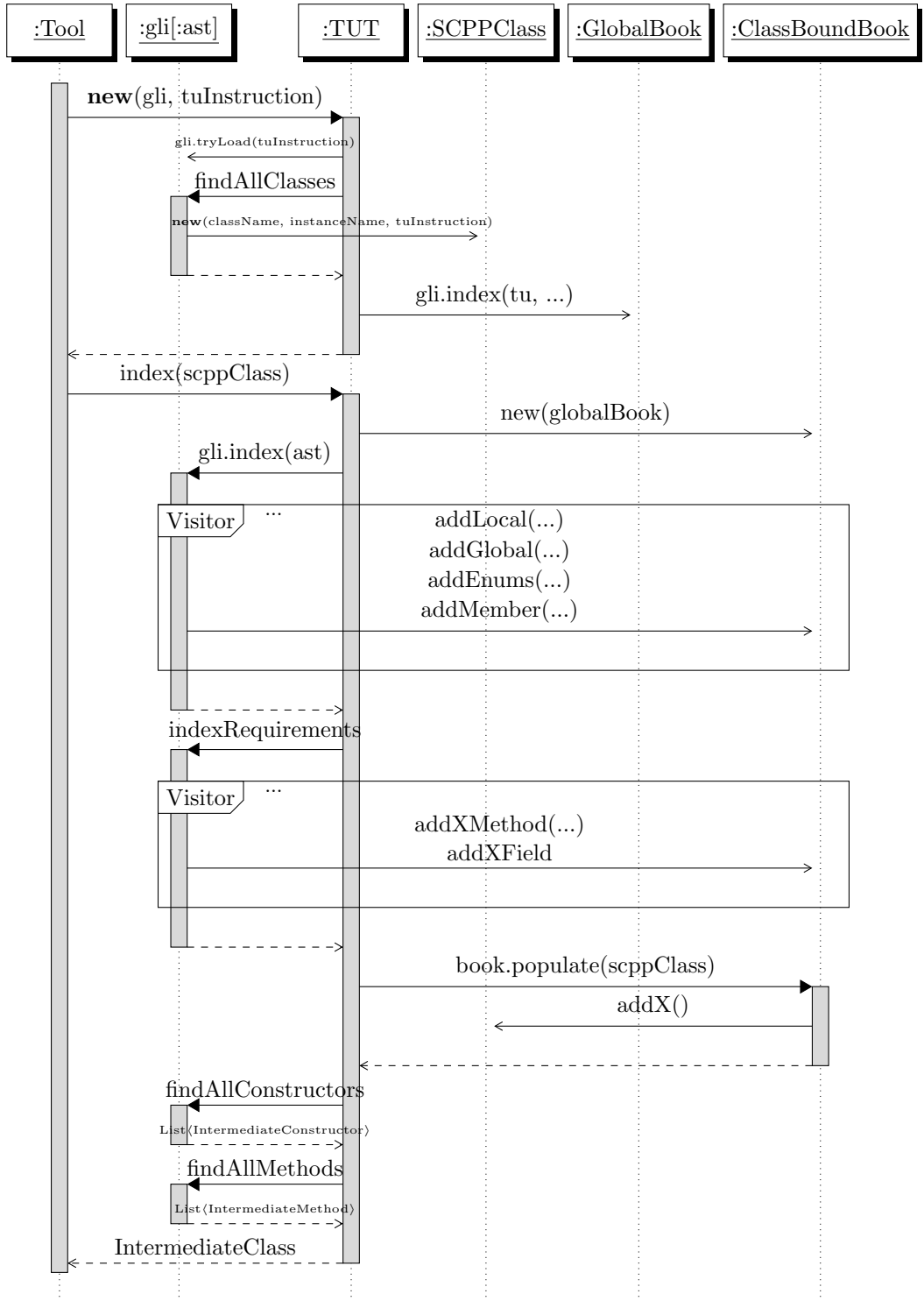

    \centering
    \begin{sequencediagram}
        \newthread{tool}{:Tool}
        \newinst{ast}{:gli[:ast]}
        \newinst[1]{tut}{:TUT}
        \newinst{class}{:SCPPClass}
        \newinst{gbook}{:GlobalBook}
        \newinst{book}{:ClassBoundBook}

        \begin{call}{tool}{\textbf{new}(gli, tuInstruction)}{tut}{}
            \mess{tut}{\tiny{gli.tryLoad(tuInstruction)}}{ast}
            \begin{call}{tut}{findAllClasses}{ast}{}
                \mess{ast}{\tiny{\textbf{new}(className, instanceName, tuInstruction)}}{class}
            \end{call}
            \mess{tut}{gli.index(tu, ...)}{gbook}
        \end{call}

        \begin{call}{tool}{index(scppClass)}{tut}{IntermediateClass}
            \mess{tut}{new(globalBook)}{book}
            \begin{call}{tut}{gli.index(ast)}{ast}{}
                \begin{sdblock}{Visitor}{...}
                \postlevel
                \mess{ast}{\shortstack{addLocal(...)\\ addGlobal(...) \\ addEnums(...)\\ addMember(...)}}{book}
                \end{sdblock}
            \end{call}

            \begin{call}{tut}{indexRequirements}{ast}{}
                \begin{sdblock}{Visitor}{...}
                \mess{ast}{\shortstack{addXMethod(...)\\ addXField}}{book}
                \end{sdblock}
            \end{call}

            \begin{call}{tut}{book.populate(scppClass)}{book}{}
                \mess{book}{addX()}{class}
            \end{call}

            \begin{call}{tut}{findAllConstructors}{ast}{\tiny{List$\langle$IntermediateConstructor$\rangle$}}
            \end{call}
            
            \begin{call}{tut}{findAllMethods}{ast}{\tiny{List$\langle$IntermediateMethod$\rangle$}}
            \end{call}
        \end{call}

    \end{sequencediagram}
    \caption{A Sequence Diagram of the transformation process (1/2) showcasing the index step.}
    \label{fig:transformation-step1}
\end{figure}

\begin{figure}
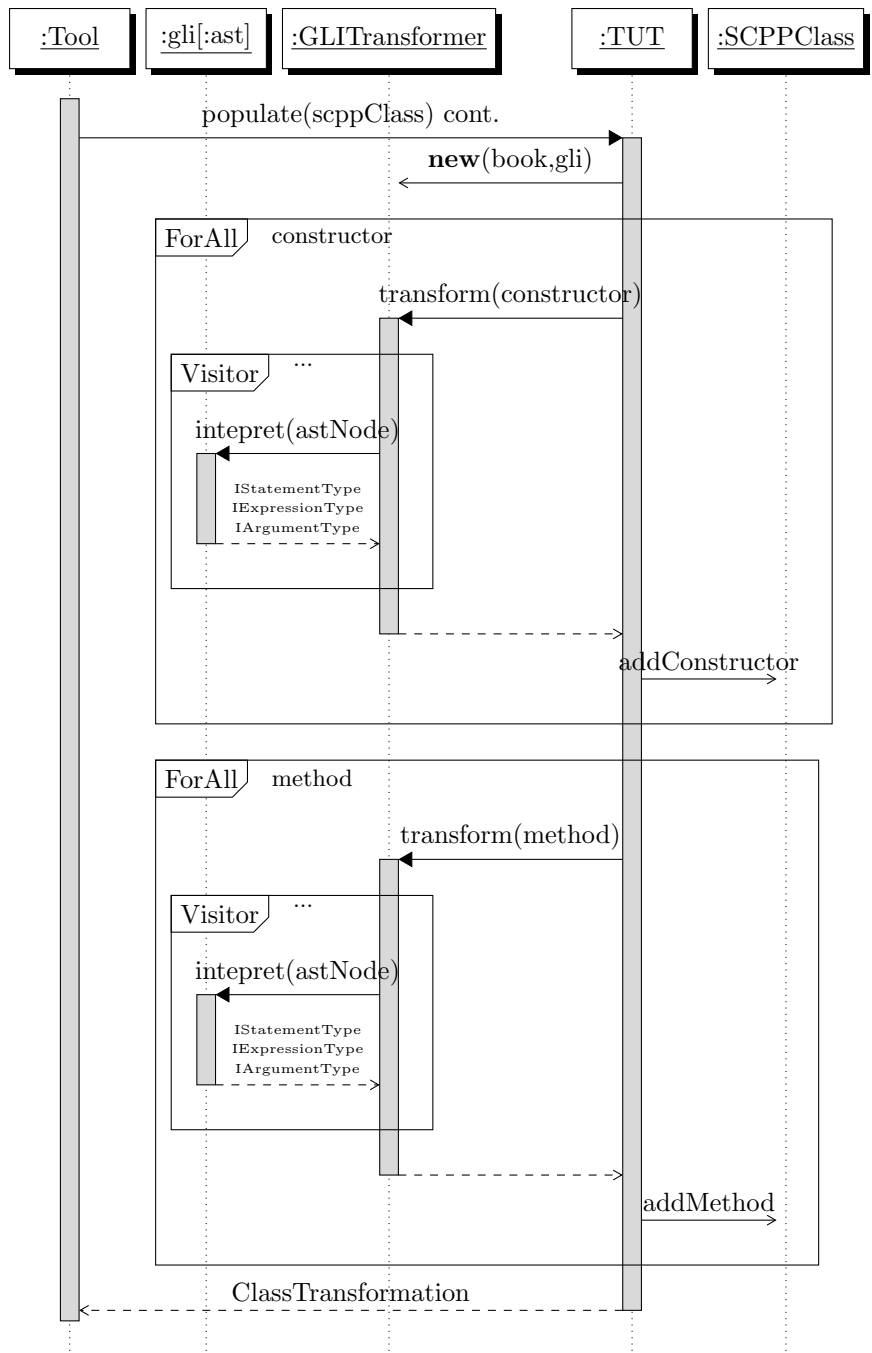

    \centering
    \begin{sequencediagram}
        \newthread{tool}{:Tool}
        \newinst{ast}{:gli[:ast]}
        \newinst{transformer}{:GLITransformer}
        \newinst[1]{tut}{:TUT}
        \newinst{class}{:SCPPClass}
        \begin{call}{tool}{populate(scppClass) cont.}{tut}{ClassTransformation}
            \mess{tut}{\textbf{new}(book,gli)}{transformer}
            \begin{sdblock}{ForAll}{constructor}
            \begin{call}{tut}{transform(constructor)}{transformer}{}
                \begin{sdblock}{Visitor}{...}
                    \begin{call}{transformer}{intepret(astNode)}{ast}{\shortstack{\tiny{IStatementType} \\ \tiny{IExpressionType} \\ \tiny{IArgumentType}}}
                        \postlevel
                    \end{call}
                \end{sdblock}
            \end{call}
            \mess{tut}{addConstructor}{class}
            \end{sdblock}

            \begin{sdblock}{ForAll}{method}
            \begin{call}{tut}{transform(method)}{transformer}{}
                \begin{sdblock}{Visitor}{...}
                    \begin{call}{transformer}{intepret(astNode)}{ast}{\shortstack{\tiny{IStatementType} \\ \tiny{IExpressionType} \\ \tiny{IArgumentType}}}
                        \postlevel
                    \end{call}
                \end{sdblock}
            \end{call}
            \mess{tut}{addMethod}{class}
            \end{sdblock}
        \end{call}
    \end{sequencediagram}
    \caption{A Sequence Diagram of the transformation process (2/2) showcasing the transfrormation step.}
    \label{fig:transformation-step2}
\end{figure}

In figures \ref{fig:transformation-step1}, and \ref{fig:transformation-step2}, we outline the transformation process of translation unit to SCPPClass in detail.
The process can be split into 2 steps: indexing (Figure \ref{fig:transformation-step1}), and transformation (Figure \ref{fig:transformation-step2}).
In the indexing step, the translation unit is traversed to index any relevant definitions and any class definition and its method and constructor definitions are stored in an intermediate and more generic data structure.
In the transformation step, the intermediate data structures are transformed into the data structures by applying the transformation rules described in the main paper.
The transformation results are then added to the SCPPClass.

We note that object-oriented programming languages mostly differ in notation, as such defining a specific transformation process for each individual language is inefficient/cumbersome.
Instead, finding a commonality between languages, and providing language-specific support only for language-specific features, should allow for the SSTraGen tool and its underlying framework to be transferable to various programming languages.
To that extent, both steps of the transformation process make use of a so-called Generic Language Interface (GLI) registration.
The GLI registration is a language-specific wrapper for the AST inside of the translation unit, and has 2 functions: the indexer, and interpreter.

The indexer is responsible for mapping language-specific constructs to one of the following generic constructs:
\begin{itemize}
    \item A variable/field declaration $\textit{Scope} \times \textit{ID}_x$ consisting of the scope and id;
    \item An Enum declaration, which consists of an enum name and 0 or more enumeratorNames;
    \item A member declaration, which consists of a class name and an instance name;
    \item Member/dynamic requirements, which track all field accesses, consisting of a ValueType and identifier, and method calls, consisting of the method signature, i.e. returnType, name, and arguments, to both members as well as any instances of classes that are passed as an argument of are created with a limited lifetime during a function call.
\end{itemize}
The scope of the variable/field declarations is used to automatically construct the scope mapping $\mathcal{S}$ used during transformation.
The requirements are used to assist with constructing bottom stubs and abstractions for members, as well as dynamic requirements.

The interpreter is responsible for mapping the language-specific ast nodes for statements, expressions, and arguments to a more generic statement/expression type.
This way the more complex transformation rules defined in Definitions \ref{def:TransS}, \ref{def:TransE}, and \ref{def:TransA} can be reused regardless of the language of the source code.

We now discuss the process described in Figure \ref{fig:transformation-step1} in further detail.
First a TranslationUnitTransformer (TUT) is constructed, during construction, the translation unit is passed to find any class definitions and any other declarations that are declared in the global namespace, such as enums.
For every class definition encountered an empty SCPPClass is constructed, i.e.\ an instance without any fields or methods.
After this, a subset of all found classes is selected for further indexing and transformation.
For each transformation target, a so-called ClassBoundBook, containing the aforementioned generic constructs, is constructed by traversing the ast associated with the said class.
During the first pass, the \textit{gli.index(ast)} call, all fields, variables, enum declarations, and member declarations are indexed.
During the second pass, the \textit{gli.indexRequirements} call, the member/dynamic requirements are indexed.
Information acquired during the first pass is sometimes required to distinguish whether a field or method call is associated with a member or a 'dynamic', hence the need for separation.
After indexing the updated ClassBoundBook is used to add fields, variables, enums, members, and requirements to the SCPPClass.
After this the only thing missing are the actual methods and constructors.

To add methods and constructors to our class, the ast node is traversed again to find all potential constructs and methods.
Any constructor and method definition is put into the more generic IntermediateConstructor and IntermediateMethod data types:
\begin{itemize}
    \item $\textbf{IConstr}(\textit{paramDecls}, \textit{initializers}, \textit{body})$, here \textit{paramDecls} is a list of all parameter declarations used for the constructor, \textit{initializers} is a list of constructor initializers, i.e.\ the $: \textit{id}(x), \hdots$ language feature in \cpp, and \textit{body} is the body of the constructor;
    \item $\textbf{IMethod}(\textit{root}, \textit{returnTy}, \textit{id}_f, \textit{paramDecls}, \textit{body})$, where \textit{root} is the ASTNode at which the method is defined, \textit{returnTy} is the return type, $\textit{id}_f$ is the method's name/identifier, and \textit{paramDecls} and \textit{body} are the same as with \textbf{IConstr}.
\end{itemize}
In both cases, the parameter declaration consists of the type, the identifier/parameter-id, and the specific name of the type in the case that the type is a PType or EnumType. In such cases, the specific type name can be used to automatically configure the associated bound.

Once all intermediate constructors and methods have been constructed, step 2 of the transformation process shown in Figure \ref{fig:transformation-step2} commences.
First, the GLITransformer is constructed.
This class is used to transform the intermediate constructors and methods by recursively applying the interpreter on the body to get the generalized Statement-, Expression-, or ArgumentType and applying the associated transformation rules outlined in the main text.
Transforming constructors is a specialized case of transforming methods, as any potential constructor initializers need to be transformed as well.
Constructor initializers are transformed as if they were assignment operators at the start of the constructor body.
Constructor initializers for members are ignored.

During the transformation process, it is possible for exceptions to be thrown, e.g.\ the interpreter has no rule for a specific AST node.
In such a case, the transformation of the specific constructor/method is interrupted, and an exception linked to the problem node is stored in a list.
Only constructors/methods that did not cause an exception during transformation are added to the SCPPClass.
If one or more exceptions were thrown as a result of the transformation process, the AST of the transformed class can be inspected, and any problem node with the associated exception is highlighted.

\begin{figure}
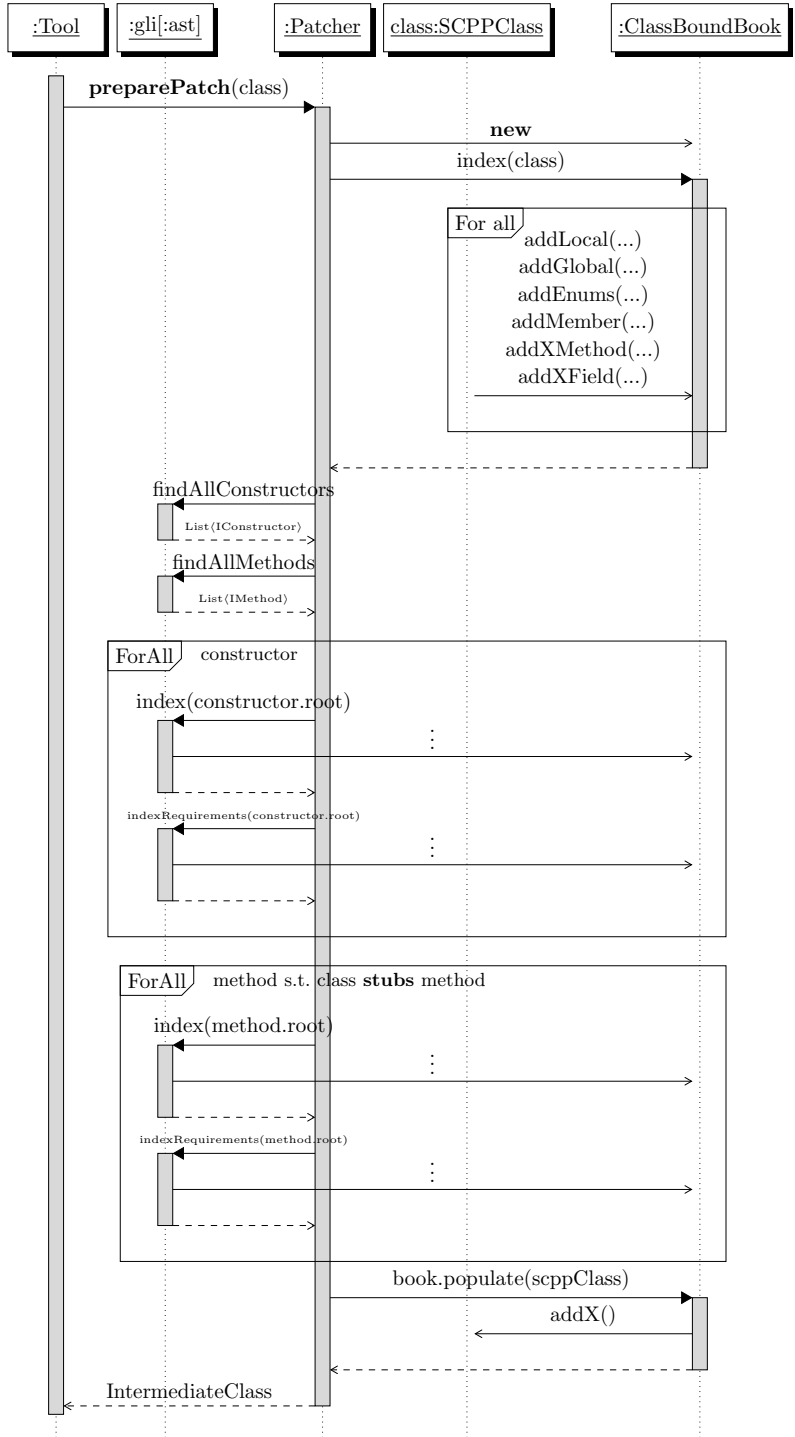

    \centering
    \scalebox{0.8}{%
    \begin{sequencediagram}
        \newthread{tool}{:Tool}
        \newinst{ast}{:gli[:ast]}
        \newinst[1]{tut}{:Patcher}
        \newinst{class}{class:SCPPClass}
        %\newinst{gbook}{:GlobalBook}
        \newinst[1]{book}{:ClassBoundBook}

        \begin{call}{tool}{\textbf{preparePatch}(class)}{tut}{IntermediateClass}
            \mess{tut}{\textbf{new}}{book}
            \begin{call}{tut}{index(class)}{book}{}
                \begin{sdblock}{For all}{}
                \postlevel \postlevel \postlevel
                \mess{class}{\shortstack{addLocal(...)\\ addGlobal(...) \\ addEnums(...)\\ addMember(...) \\ addXMethod(...) \\ addXField(...)}}{book}
                \end{sdblock}
            \end{call}

            \begin{call}{tut}{findAllConstructors}{ast}{\tiny{List$\langle$IConstructor$\rangle$}}
            \end{call}
            
            \begin{call}{tut}{findAllMethods}{ast}{\tiny{List$\langle$IMethod$\rangle$}}
            \end{call}

            \begin{sdblock}{ForAll}{constructor}
            \begin{call}{tut}{index(constructor.root)}{ast}{}
                \mess{ast}{$\vdots$}{book}
            \end{call}
            \begin{call}{tut}{\tiny{indexRequirements(constructor.root)}}{ast}{}
                \mess{ast}{$\vdots$}{book}
            \end{call}
            \end{sdblock}

            \begin{sdblock}{ForAll}{method s.t.\ class \textbf{stubs} method}
            \begin{call}{tut}{index(method.root)}{ast}{}
                \mess{ast}{$\vdots$}{book}
            \end{call}
            \begin{call}{tut}{\tiny{indexRequirements(method.root)}}{ast}{}
                \mess{ast}{$\vdots$}{book}
            \end{call}
            \end{sdblock}

            \begin{call}{tut}{book.populate(scppClass)}{book}{}
                \mess{book}{addX()}{class}
            \end{call}
            
        \end{call}
    \end{sequencediagram}}
    \caption{A Sequence Diagram showcasing the patch step replacing the standard indexing step.}
    \label{fig:transformation-patch}
\end{figure}

In languages such as \cpp, it is possible to split up the declaration and definitions of a class into multiple files.
In such cases, the AST node of the class definition is not an (indirect)parent of the AST nodes of the function bodies/constructors, and as such the standard index process will not capture local information such as variables and dynamic requirements.
To solve this, a special version of the indexing step exists, the so-called Patch step, outlined in Figure \ref{fig:transformation-patch}.
The transformation for a multi-file class thus takes place as follows: For the main file, i.e.\ the file with the class definition, step 1, followed by step 2 are executed. 
Then for any additional file, the \textit{preparePatch} step followed by step 2 is executed.
The \textit{ScppClass} instance in the \textit{preprarePatch} step is the scppClass constructed from the main file in step 1.

At the start of the \textit{preparePatch} step, all variable, fields, etc.\ that were previously found and registered with the SCPPClass are added to a new ClassBoundBook.
This operation is effectively the inverse of the \textit{book.populate(scppClass)} call occurring during step 1.
After creating the new ClassBoundBook, all constructors and methods associated with the incomplete class are indexed by traversing the ast of the separate translation unit.
After this the book is indexed further using any variable declarations and requirements found within the root ast's of the new methods and constructors.
Finally, the scppClass is updated using the updated ClassBoundBook, after which the populate function is executed, which will iterate/transform the newly found constructors and methods, as per step 2.

\end{appendix}
\end{document}